\documentclass{article}

\usepackage{microtype}
\usepackage{graphicx}
\usepackage{subcaption}
\usepackage{booktabs} %

\usepackage{hyperref}

\usepackage[preprint]{icml2026}

\usepackage{amsmath}
\usepackage{amssymb}
\usepackage{mathtools}
\usepackage{amsthm}

\usepackage[capitalize,noabbrev]{cleveref}

\usepackage{xspace}
\usepackage{xcolor}
\usepackage{colortbl} %
\usepackage{bbm}
\usepackage{listings} %
\usepackage{tcolorbox} %

\usepackage{booktabs} %
\usepackage{multirow} %
\usepackage{pgfplots}
\usepackage{tikz}
\usepackage{xcolor}
\usepackage{float}
\usepackage{wrapfig}
\usepackage{enumitem}
\usepackage{bm}
\usepackage{arydshln}

\theoremstyle{plain}

\theoremstyle{definition}

\theoremstyle{remark}

\usepackage[textsize=tiny]{todonotes}

\newcommand{\ours}{\textsc{TritonRL}\xspace}

\icmltitlerunning{TritonRL: Training LLMs to Think and Code Triton Without Cheating}

\begin{document}

\twocolumn[
  \icmltitle{TritonRL: Training LLMs to Think and Code Triton Without Cheating}

  \icmlsetsymbol{wda}{*}
  \icmlsetsymbol{wda2}{\ddag}
  \icmlsetsymbol{cor}{\dag}

  \begin{icmlauthorlist}
    \icmlauthor{Jiin Woo}{yyy,wda,cor}
    \icmlauthor{Shaowei Zhu}{comp}
    \icmlauthor{Allen Nie}{comp2,wda2}
    \icmlauthor{Zhen Jia}{comp}
    \icmlauthor{Yida Wang}{comp}
    \icmlauthor{Youngsuk Park}{comp,cor}
  \end{icmlauthorlist}

  \icmlaffiliation{yyy}{Carnegie Mellon University}
  \icmlaffiliation{comp}{Amazon Web Services}
  \icmlaffiliation{comp2}{Google DeepMind}

  \icmlcorrespondingauthor{Jiin Woo}{jiinw@andrew.cmu.edu}
  \icmlcorrespondingauthor{Youngsuk Park}{pyoungsu@amazon.com}

  \icmlkeywords{Machine Learning, ICML}

  \vskip 0.3in
]

\printAffiliationsAndNotice{\workdoneawsintern \workdoneaws}  %

\begin{abstract}
The rapid evolution of Large Language Models (LLMs) has driven a growing demand for automated, high-performance system kernels to accelerate machine learning workloads. We introduce \ours, a domain-specialized 8B-scale LLM for Triton programming, trained via a novel reinforcement learning (RL) framework. While Triton synthesis faces unique challenges, including data scarcity and a high susceptibility to reward hacking, our approach enables robust kernel generation through two primary innovations. First, we implement a multi-layered verification system that provides high-fidelity reward signals, ensuring that generated kernels are both syntactically and functionally valid. Second, we propose Hierarchical Reward Decomposition (HRD), which decouples reinforcement for high-level reasoning and low-level implementation to resolve the credit assignment problem in long-sequence generation.
Comprehensive evaluations on KernelBench demonstrate that \ours achieves state-of-the-art correctness and runtime speedup, outperforming concurrent Triton-specific models and matching the performance of frontier models with over 100B parameters. Our results highlight the effectiveness of hardware-aware RL paradigms in specialized domain adaptation.
\end{abstract}

\section{Introduction}
The exponential growth in demand for GPU computing resources has driven the need for highly optimized GPU kernels that improve computational efficiency, yet with the emergence of numerous GPU variants featuring diverse hardware specifications and the corresponding variety of optimization kernels required for each, developing optimized kernels has become an extremely time-consuming and challenging task. In response to this need, there is growing interest in leveraging large language models (LLMs) for automated kernel generation. While there have been attempts introducing inference frameworks that utilize general-purpose models, such as OpenAI models and DeepSeek, for generating kernels \citep{ouyang2025kernelbench,lange2025ai,nvidia2025deepseek_kernelgen, muhamed2023training}, they often struggle with even basic kernel implementations, thereby highlighting the critical need for domain-specific models specifically tailored for kernel synthesis.

As the need for specialized models for kernel generation has emerged, several works have focused on fine-tuning LLMs for CUDA or Triton. In the CUDA domain, recent RL-based approaches include Kevin-32B \citet{baronio2025multi}, which progressively improves kernels using execution feedback as reward signals, and CUDA-L1, which applies contrastive RL to DeepSeek-V3 \citet{li2025cudal1}. While these large models (32B-671B parameters) achieve strong CUDA performance, their training costs remain prohibitively expensive. 
Other research has focused on Triton, a domain-specific language that provides a higher-level abstraction than CUDA for writing efficient GPU kernels. Unlike CUDA, which benefits from decades of development and abundant training data, Triton is newer and has fewer high-quality kernel examples, making it more difficult to train specialized models. Recent works have trained LLMs for Triton kernel generation via supervised fine-tuning (SFT) on torch compiler–generated code \citep{kernelllm2025} or via LLM distillation followed by RL with execution feedback \citep{li2025autotriton}, typically at the 8B parameter scale. Although these smaller models improve over their base models, there remains substantial room to improve efficiency and correctness.

Furthermore, there is a common issue reported across kernel generation works, reward hacking \citep{baronio2025multi,li2025autotriton}. Due to the scarcity of high-quality kernel examples compared to other programming languages, most approaches rely on RL training using runtime measurements and correctness rewards from unit tests after kernel execution. However, models frequently learn to exploit unit test loopholes, such as direct use of high-level PyTorch modules, rather than generating proper code, and this phenomenon is particularly prevalent in smaller models (8B and below) \citep{baronio2025multi}. This issue fundamentally undermines the core objective of developing more efficient custom kernels to replace existing pre-optimized libraries, while current approaches predominantly rely on simple rule-based syntax verification whose effectiveness remains uncertain.

In this paper, we present \ours, an 8B-scale LLM specialized for Triton programming that achieves state-of-the-art performance in both functional correctness and execution speed. Our training pipeline enables 8B-scale models to robustly surpass much larger frontiers through the following contributions:

\begin{itemize}[leftmargin=.5cm]
 \item \textbf{Robust Verification Framework}: We introduce a multi-layered verification system designed specifically to mitigate Triton-specific reward hacking. By identifying failure modes where models bypass core custom kernel logic, we implement fine-grained verifiers, utilizing both rule-based heuristics and LLM-based judges, that ensure reward signals are grounded in genuine functional and syntactic integrity.
 
 \item \textbf{Hierarchical Reward Decomposition (HRD)}: We propose a novel RL optimization objective that addresses the credit assignment problem in long-sequence generation. By decoupling rewards for high-level reasoning (planning) and low-level implementation (coding), we provide targeted reinforcement for strategy and execution independently. This approach outperforms uniform sequence-level rewards in both accuracy and speedup.

\item \textbf{Comprehensive Evaluation and State-of-the-Art Performance}: We provide a rigorous evaluation of open-source and proprietary models using our verification framework, revealing critical performance gaps in functional validity. Our results demonstrate that \ours sets a new state-of-the-art for 8B-scale models and performs comparably to or better than frontier models with over $100$B parameters.\footnote{Codes and datasets will be available soon.} %
\end{itemize}

\subsection{Related Work}
\paragraph{Large Language Models for Kernel Generation}
The demand for GPU efficiency has spurred interest in using large language models (LLMs) for automated kernel generation in CUDA and Triton \citep{ouyang2025kernelbench,li2025tritonbench,nvidia2025deepseek_kernelgen}. Building on this, benchmarks like KernelBench \citep{ouyang2025kernelbench} and TritonBench \citep{li2025tritonbench} have been introduced to evaluate LLMs on kernel optimization. Although LLMs show promise in general-purpose programming, they often achieve low success rates on specialized GPU programming tasks \citep{ouyang2025kernelbench}, highlighting the need for domain-specific models tailored to kernel synthesis.

Recent research focuses on domain-specific fine-tuning. For CUDA, models like Kevin-32B \citep{baronio2025multi} and CUDA-L1 \citep{li2025cudal1} utilize multi-turn or contrastive RL to optimize performance. In the Triton domain, KernelLLM \citep{kernelllm2025} performs SFT on the KernelBook dataset \citep{kernelbook2025}, while AutoTriton \citep{li2025autotriton} applies SFT and RL with verifiable rewards based on correctness.
Both KernelLLM and AutoTriton are concurrent works developed alongside our work, and we provide a detailed performance comparison in Section~\ref{sec:main_results}.

\paragraph{Reinforcement Learning with Verifiable Rewards}
Reinforcement Learning with Verifiable Rewards (RLVR) has become a standard for training LLMs in objective domains like mathematics and coding \citep{shao2024deepseekmath,guo2025deepseek, AlphaCode2022}. In these frameworks, the accuracy of the reward signal is critical to prevent reward hacking \citep{skalse2022defining}, where models exploit test suites without genuinely solving the problem \citep{sharma2024critical,gao2024designing}.
In kernel generation, reward hacking typically involves models using high-level PyTorch modules to bypass custom kernel implementation while still passing unit tests. While some recent efforts \citep{li2025autotriton, baronio2025multi} have introduced syntax checks, these checks remain insufficient to detect subtler forms of reward hacking such as computation delegation, leaving a critical gap in reward fidelity.

\begin{figure*}[t]
  \centering
  \includegraphics[width=0.95 \linewidth]{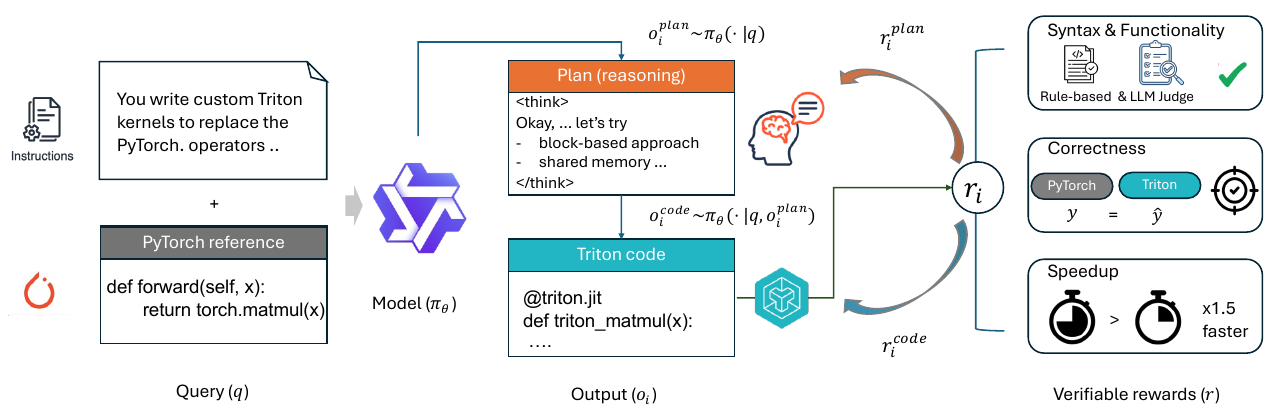}
  \caption{RL training workflow of \ours. The framework illustrates how kernel generation flows through fine-grained verification layers and hierarchical reward decomposition, providing robust and targeted feedback separately to reasoning traces (plan tokens) and to Triton code (code tokens) conditioned on those plans.}
  \label{fig:plancode_example}
  \vspace{-2mm}
\end{figure*}

\section{\ours}
In this section, we introduce \ours, a comprehensive reinforcement learning (RL) framework designed to optimize language models for high performance Triton kernel generation. Our approach addresses the dual challenge of achieving functional correctness and hardware specific efficiency, optimizing for execution latency on target accelerators. To this end, our methodology is built upon three core components: (1) base model priming and task curation, (2) reward formulation with robust verification, and (3) policy optimization with hierarchical reward decomposition. %

\subsection{Base Model Priming and Task Curation}
For a model to effectively optimize Triton kernels via RL, it must first possess a foundational understanding of Triton syntax and be exposed to a high quality task distribution during RL. We address these requirements through two efforts: (1) base model priming via knowledge distillation to instill fundamental programming skills and (2) task curation via data augmentation and difficulty aware data mixing to construct an optimal training distribution for RL.

\subsubsection{Triton Knowledge Distillation via SFT}
\label{sec:sft}
Recent studies indicate that pretrained language models, particularly at the 8B scale, exhibit limited proficiency in Triton programming, often struggling with low level syntax and hardware specific optimization \citep{kernelllm2025,li2025autotriton}. To bridge the gap, we perform supervised fine-tuning (SFT) on a base model to distill expertise from frontier models into a more efficient base. In this paper, we utilize Qwen3-8B \citep{qwen3tech} as our base model.

Our primary data source is KernelBook \citep{kernelbook2025}, which contains a rich collection of Triton kernel generation tasks paired with reference PyTorch implementations. We define $\mathcal{G}_{KB} = \{g_i\}_{i \in \{1, \dots, 11K\}}$ as a set of 11K tasks sampled from KernelBook. For each task query $q_i = q(g_i)$, wrapped with an instruction to generate a Triton kernel for task $g_i$, we prompt a teacher model to generate Triton kernels multiple times, denoted by $m_i \in [1, 10]$. Each resulting output $o_{i,j}$ (where $j \in \{1, \dots, m_i\}$) consists of a synthesized reasoning trace followed by the corresponding Triton implementation. For a given teacher model $\pi_{\text{teacher}}$, this process yields an SFT dataset:$$\mathcal{D}_{SFT}= \big\{(q_i, o_{i,j}) \big\}_{g_i \in \mathcal{G}_{KB}, q_i = q(g_i), \{o_{i,j}\}_{j=1}^{m_i} \sim \pi_{\text{teacher}}(\cdot|q_i)}.$$ 

With the SFT dataset, we train our base model parameterized by $\theta$, $\pi_{\theta}$, to maximize the log-likelihood:
\begin{equation*}
\max_{\theta} \mathbb{E}_{(q_i, o_{i,j}) \sim \mathcal{D}_{SFT}} \left[ \log \pi_{\theta}(o_{i,j}|q_i) \right].
\end{equation*}
Through SFT, it learns to internalize the underlying logic of kernel optimization rather than simply memorizing syntax. To explore the impact of different teacher expertise, we construct two separate SFT datasets (approximately 60K samples each) using DeepSeek R1 \citep{guo2025deepseek} and GPT OSS 120B \citep{openai2025gptoss120b} respectively, resulting in two distinct SFT models. Examples of task queries and outputs are provided in Appendix \ref{sec:kernelbook_example}, with additional training details in Appendix \ref{sec:appendix:hps}.

\subsubsection{Data Augmentation and Difficulty-Aware Data Mixing}
\label{sec:rl_data_construction}

The selection of training data is crucial for effective RL post-training, and many works have shown that selective sampling by leveraging additional information, such as difficulty or interaction between data points, can significantly improve model performance \citep{yu2025dapo,chen2025acereason,chen2024aioli}.
While KernelBook provides a large-scale and diverse set of kernels, naive uniform sampling may lead to suboptimal efficiency. Consequently, we augment each task and apply a difficulty aware mixing strategy to refine the RL training set.

\paragraph{Diverse Input Augmentation.} We start from the KernelBook task set $\mathcal{G}_{KB} = \{g_i\}_{i \in \{1, \dots, 11K\}}$, used to construct the SFT datasets. To increase diversity and robustness, we augment each task $g_i$ with additional input-generating functions using GPT-OSS 120B, producing multiple tasks $\{\tilde{g}_{i,j}\}_{j \in \{1, \dots, n_i\}}$ with $n_i$ different input shapes. After augmentation, we validate the tasks by execution, so the number of valid tasks $n_i \le 5$ may vary.
We denote the resulting set of augmented tasks as $\tilde{\mathcal{G}}_{KB} = \{\tilde{g}_{i,j} \}_{i \in \{1, \dots, |\mathcal{G}_{KB}| \}, j \in \{1, \dots, \tilde{m}_i\}}$. By default we use the original task set $\mathcal{G}_{KB}$; when RL training uses the augmented set $\tilde{\mathcal{G}}_{KB}$ we indicate this with “+ IA”. In Section~\ref{sec:main_results}, we analyze the impact of input augmentation on final performance.

\paragraph{Difficulty-Aware Data Mixing.}
Then, following the difficulty levels defined in KernelBench \citep{ouyang2025kernelbench},  we classify tasks into two levels, where Level 1 represents single-kernel tasks, such as convolution, and Level 2 represents fusion tasks, such as conv+bias+ReLU. %
To label each task $g_i$, we utilize an LLM based labeler (Qwen3-235B-Instruct \citep{qwen3tech}) and let the difficulty label as $d_i = d(g_i) \in \{1, 2\}$. Focusing on Level 1 and Level 2 tasks, we denote the subset of tasks at difficulty level $d$ as $\mathcal{G}_{KB}^{(d)} = \{g_i \in \mathcal{G}_{KB} | d(g_i) = d\}$ for $d \in \{1, 2\}$ and construct RL training datasets according to various mixture probabilities $\bm{p}$ as follows:
$$
\mathcal{D}_{RL}(\bm{p}) = \{(g_i, q_i)\}_{d \sim \bm{p}, g_i \sim \mathcal{G}_{KB}^{(d)}, q_i = q(g_i)},
$$
where $\bm{p} = (p^{(1)}, p^{(2)}) \in [0,1]^2$ where $p^{(d)}$ denotes the sampling probabilities for difficulty level $d$. In our experiments (Section~\ref{sec:main_results}), we explore different training data mixtures and evaluate their effectiveness during RL post-training to identify the optimal configuration.
Throughout this paper, we set $\bm{p} = (1.0, 0.0)$ unless otherwise specified and omit $\bm{p}$ for notational simplicity.
Additional details regarding the labeling prompt is provided in Appendix \ref{sec:appendix:data-mixing-subset}.

Building on the model equipped with essential Triton skills, $\pi_{\theta}$ and a curated RL dataset $\mathcal{D}_{RL}$, we now focus on the core of our RL framework.
Before diving into the specifics, we outline the overall RL training workflow of \ours. For each task $(g_i, q_i)$ sampled from the RL training dataset $\mathcal{D}_{RL}$, the model generates a group of outputs $\{o_{i,j}\}_{j=1}^G$ for group size $G$, where each output $o_{i,j} = \{o_{i,j}^\text{plan}, o_{i,j}^\text{code}\}$ consists of reasoning traces planning on how to optimize Triton kernels ($o_{i,j}^\text{plan}$) followed by Triton code ($o_{i,j}^\text{code}$).
After executing the generated Triton code, the model receives reward feedback $r_{i,j}$ based on the quality of the output, which it uses to update its parameters $\theta$ to maximize expected rewards over time. We illustrate the overall RL training workflow of \ours in Figure~\ref{fig:plancode_example}.

\begin{figure}[t!]
  \centering
  \includegraphics[width=\columnwidth]{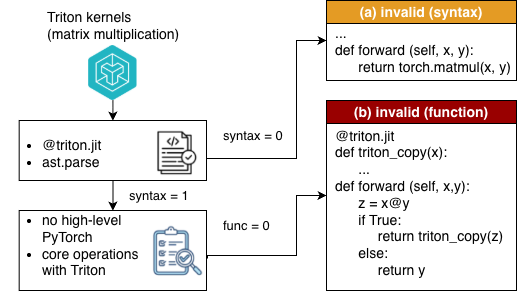}
  \caption{Illustration of the flow of our robust verifier incorporating syntax and functionality checkers and the examples of invalid Triton codes. (a) invalid syntax: the code lacks any Triton blocks with @triton.jit. 
  (b) invalid functionality: the code include dummy Triton code that just copies data without meaningful operation delegating core operation (matrix multiplication) to PyTorch modules (torch.matmul).}
  \label{fig:invalid_example}
  \vspace{-5mm}
\end{figure}

\subsection{Reward Formulation with Robust Verification}
In reinforcement learning, reward design is critical, as poorly aligned signals can lead to reward hacking, where the model exploits loopholes rather than achieving the intended objectives. We identify specific Triton failure modes, such as ignoring core kernel functionality, that are often missed by conventional reward designs based solely on unit tests. To mitigate this, we develop fine grained verification layers that rigorously assess code quality across multiple dimensions.

For a generated output $o$ and a corresponding reference PyTorch code $g$, we introduce verifiers that evaluate the following aspects:
\begin{itemize}[leftmargin=*]
\item $\texttt{valid}$: A binary verifier to check output code is syntactically and functionally valid Triton code. It is defined as $\texttt{valid}(o) = \texttt{syntax}(o) \cdot \texttt{func}(o)$, where:
\begin{itemize}[leftmargin=.5cm]
    \item 
    $\texttt{syntax}$
    : A binary verifier that assesses whether code $o^\text{code}$ is valid Triton syntax. We use a rule-based linter to verify the presence of Triton kernels annotated with @triton.jit. 
\item $\texttt{func}$: 
A binary functionality verifier to detect whether $o^\text{code}$ constitutes a valid Triton kernel. Syntax checks alone are insufficient, since models may output code that superficially passes verification but defers computations to high-level PyTorch modules (e.g., torch.nn, @) or hardcodes constants, as in Figure~\ref{fig:invalid_example} (b). To address this, we combine a rule-based linter, which detects actual calls of Triton kernels and flags reliance on PyTorch modules, combined with an LLM-based judge (Qwen3-235B-Instruct~\citep{qwen3tech}) that evaluates semantic correctness against task specifications.
In Appendix~\ref{sec:appendix:cheating_detector}, we provide additional details on the LLM judge used for functionality verification.
\end{itemize}
See Figure~\ref{fig:invalid_example} for the illustration of invalid Triton codes that fail syntax and functionality checks.

\item $\texttt{compiled}$: A binary verifier that checks whether a Triton code can be successfully compiled without errors.

\item $\texttt{correct}$: A binary verifier that evaluates whether a Triton code produces correct outputs by compiling and comparing its results against those of the reference code $g$ for five random test inputs $\{x_l\}_{l=1}^{5}$.
\begin{equation*}
\scalebox{0.85}{ 
  $\displaystyle 
  \begin{aligned}
\texttt{correct}(g, o) \;&=\; \texttt{compiled}\!\left(o\right) \cdot 
\prod_{l=1}^{5} \mathbbm{1}\!\left[\, o^{\text{code}}(x_l) == g(x_l) \,\right]
\end{aligned}
  $
}
\end{equation*}

\item $\texttt{speedup}$: A scalar score that quantifies the execution time improvement of the generated Triton code $o^\text{code}$ relative to the reference code $g$. For a test input $x$,
\begin{equation*}
\scalebox{0.85}{ 
  $\displaystyle 
  \begin{aligned}
\texttt{speedup} (g, o) = \frac{\tau(g, x)}{\tau(o^\text{code}, x)} \cdot 
\texttt{correct}(g, o),
\end{aligned}
  $
}
\end{equation*}
where $\tau(\cdot, x)$ is the average runtime of a given code for input $x$ measured over ten runs.
\end{itemize}

\begin{figure}[t!]
    \centering
    \includegraphics[width=\linewidth]{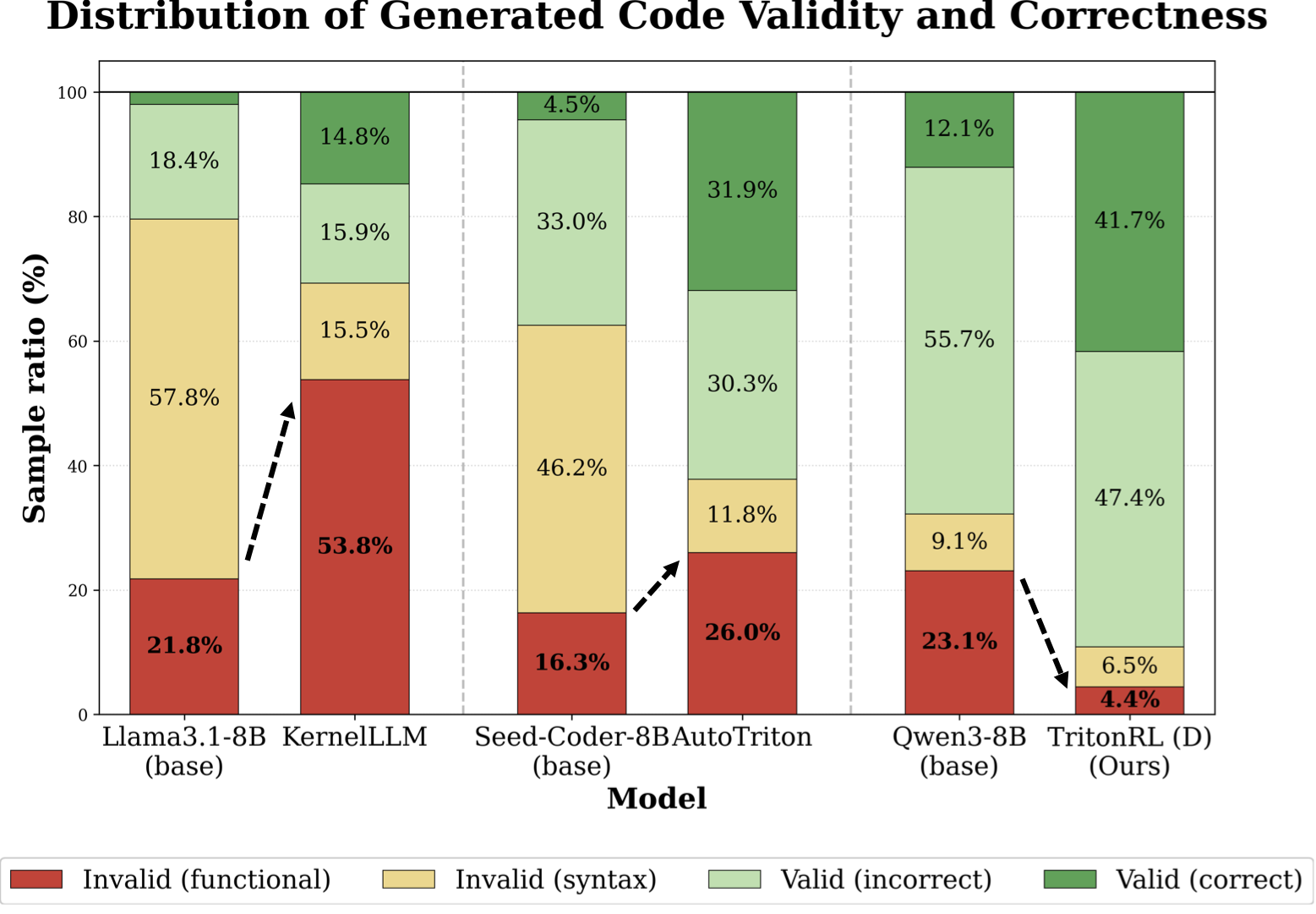}
    \caption{Side-by-side comparison of fine-tuned models for Triton programming (KernelLLM and AutoTriton) and their base models. 
    For each metric, the numbers represent the ratio of corresponding samples among 10 generated samples for each of the 100 tasks in KernelBench Level 1.  \ours (D) denotes our model trained from SFT model distilled from DeepSeek-R1. KernelLLM and AutoTriton show increased functional invalidity (red) after fine-tuning, while \ours significantly reduces such errors.}
    \label{fig:reward_hacking}
    \vspace{-3mm}
\end{figure}

In Figure~\ref{fig:reward_hacking}, we analyze the validity of Triton codes generated by existing specialized models, such as AutoTriton \citep{li2025autotriton} and KernelLLM \citep{kernelllm2025}, and observe a high frequency of functionality violations (red). Interestingly, their base models exhibit lower error rates prior to fine-tuning. This suggests that fine-tuning processes that lack rigorous functionality checks can inadvertently incentivize models to adopt invalid shortcuts, superficially satisfying syntax requirements while failing to implement the core kernel operations.

To mitigate these failure modes, we define two reward functions that deliver robust, accurate signals for correctness and efficiency as follows:
\begin{equation} \label{eq:reward_functions}
\scalebox{0.9}{ 
  $\displaystyle 
  \begin{aligned}
R^{\text{correct}}(g, o) &= \texttt{valid}(o) \cdot \texttt{correct}(g, o), \\
R^{\text{speedup}}(g, o) &= \texttt{valid}(o) \cdot \text{clip} \left(\texttt{speedup}(g, o), 2 \right).
\end{aligned}
  $
}
\end{equation}
The syntax and functionality checks both serve as necessary conditions for both reward signals to prevent the model from being incentivized to exploit shortcuts. Also, we clip the speedup at $2$ to prevent extreme values from destabilizing training.

\subsection{Policy Optimization with Hierarchical Reward Decomposition}
Building upon the robust reward designs, we introduce a policy optimization framework designed to leverage these signals effectively. While the verification layers ensure high-fidelity feedback, training language models with long reasoning traces remains challenging due to the credit assignment problem. When a single, monolithic reward is applied to a lengthy response, it fails to distinguish between the tokens that contribute to a brilliant optimization strategy and those that contain a minor syntax error in the implementation.

This issue is particularly pronounced in Triton kernel code generation, where a model may propose a promising memory orchestration plan in its reasoning trace, yet fail the entire task due to a low level indexing error in the subsequent code. Uniformly penalizing such a response conflates high quality reasoning with poor execution \citep{qu2025rlad}, preventing the model from internalizing effective optimization strategies.

\paragraph{GRPO with Hierarchical Reward Decomposition (HRD).}
To address this, we propose an optimization approach based on Group Relative Policy Optimization (GRPO) \citep{shao2024deepseekmath}, augmented with Hierarchical Reward Decomposition (HRD). GRPO is an actor-only RL algorithm that optimizes the policy by comparing the rewards of multiple outputs within a group generated for the same query, eliminating the need for a separate critic model.

Our key innovation is to segment the long output sequence into two hierarchical actions: high-level planning ($o^\text{plan}$) and low-level coding ($o^\text{code}$). By assigning distinct reward credits to these segments (Figure \ref{fig:plancode_example}), we better align the Triton kernel optimization plan with the final implementation. We decompose the objective into two components and optimize them jointly:
\begin{equation*}
\scalebox{0.95}{ 
  $\displaystyle 
  \begin{aligned}
  \mathcal{J}(\theta) &= \mathbb{E}_{q_i \sim \mathcal{D}_{RL}, ~ \{o_{i,j} \}_{j=1}^G \sim \pi_{\theta_{old}}(\cdot|q_i) } \left [  \alpha \mathcal{J}_i^{\text{\tiny plan}} (\theta) + \mathcal{J}_i^{\text{\tiny code}} (\theta) \right ]
  \end{aligned}
  $
}
\end{equation*}
where $\alpha \in [0,1]$ is a weighting factor that balances the update rates between the model's planning and coding actions.
For each action class $c \in \{\text{plan}, \text{code}\}$, the class-specific objective $\mathcal{J}_i^{c} (\theta)$ is defined as:
 \begin{equation*}
 \scalebox{0.85}{ 
   $\displaystyle 
   \begin{aligned}
  \mathcal{J}_i^{c}(\theta) = \frac{1}{G} \sum_{j=1}^G \frac{1}{|o_{i,j}|} \sum_{t\in T_{i,j}^{c}}  \min \Big [ &A_{i,j}^{c} s_{i,j,t}^{c}(\theta) , \cr
  & A_{i,j}^{c} \text{clip} \left( s_{i,j,t}^{c}(\theta) , 1-\epsilon, 1+\epsilon \right) \Big ],
  \end{aligned}
   $
 }
 \end{equation*}
where $T_{i,j}^c$ represents the set of token indices for class $c$ in output $o_{i,j}$ and the token-wise probability ratios are given by
\begin{equation*}
\scalebox{0.9}{ 
  $\displaystyle 
  \begin{aligned}
  s_{i,j,t}^{\text{\tiny plan}}(\theta) &= \frac{\pi_\theta(o_{i,j,t}^{\text{\tiny plan}}|q_i, o_{i,j,<t}^{\text{\tiny plan}})}{\pi_{\theta_{old}}(o_{i,j,t}^{\text{\tiny plan}}|q_i, o_{i,j,<t}^{\text{\tiny plan}})}, \\
  s_{i,j,t}^{\text{\tiny code}}(\theta) &= \frac{\pi_\theta(o_{i,j,t}^{\text{\tiny code}}|q_i, o_{i,j}^{\text{\tiny plan}}, o_{i,j,<t}^{\text{\tiny code}})}{\pi_{\theta_{old}}(o_{i,j,t}^{\text{\tiny code}}|q_i, o_{i,j}^{\text{\tiny plan}}, o_{i,j,<t}^{\text{\tiny code}})}.
  \end{aligned}
  $
}
\end{equation*} The group-wise advantages $A_{i,j}^c$ are computed as $A_{i,j}^c = r_{i,j}^c - \frac{1}{G} \sum_{j=1}^G r_{i,j}^c$, where we assign different reinforcement signal for each token class as follows:
\begin{align} \label{eq:reward}
r_{i,j}^{\text{\tiny plan}} = R^{\text{speedup}}(g_i, o_{i,j}), \quad
r_{i,j}^{\text{\tiny code}} = R^{\text{correct}}(g_i, o_{i,j}),
\end{align}
where the reward functions are defined in \eqref{eq:reward_functions}.

\textbf{Hierarchical vs Uniform Rewards.} 
Unlike uniform reward designs \citep{li2025autotriton, baronio2025multi} that apply the same signal to all tokens, our HRD provides targeted feedback. In \eqref{eq:reward}, we assign speedup-based rewards to plan tokens to encourage the discovery of efficient optimization strategies, while assigning correctness-based rewards to code tokens. This separation ensures that reasoning traces are encouraged to propose optimization strategies that yield efficient kernels, while code generation is guided to produce valid implementations that faithfully realize these plans.
Assigning correctness-based rewards to code tokens is deliberate because the implementation is conditioned on the plan, penalizing code for low speedup would unfairly punish a correct implementation of an inherently suboptimal plan.
Correctness rewards for code tokens decouple this confound,
allowing the code policy to be reinforced purely for faithful
realization of a given plan.
 We validate this design choice in Appendix~\ref{sec:reward_assignment_ablation}, where we compare assigning correctness-based versus speedup-based rewards to code tokens, supporting our rationale.

\textbf{Balancing Planning and Coding Updates.} 
The weighting factor $\alpha$ manages the co-evolution of the planning and coding policies. To effectively master both, the model requires a stable "practice environment." If $\alpha=1.0$, the planning distribution shifts too rapidly for the coding policy to stabilize. By setting $\alpha \approx 0.1$, we ensure the planning distribution evolves gradually, giving the coding policy sufficient "time" to learn correct implementations for a given set of plans. This avoids the premature rejection of promising optimization strategies due to transient implementation errors in early training phases.

\begin{table*}[ht!] \small
    \caption{Main results on KernelBench Level 1 and Level 2. 
    All metrics are reported as pass@10 (\%). Our model achieves the best results among models with fewer than 32B parameters. 
    The left block reports evaluation with the robust verifier (syntax + functionality). 
    The right block (w/o robust verifier) lacks functionality checks, leading to misleading correctness estimates. We denote our model trained from SFT with DeepSeek-R1 (resp. GPT-OSS 120B) as \ours (D) (resp. (G)) and report the results of SFT-only variants without RL for reference (See Section~\ref{sec:sft}). We additionally denote our models trained with input augmentation (IA) as + IA, which uses RL datasets where diverse input shapes are augmented (See Section~\ref{sec:rl_data_construction}).
    }
    \label{tab:kernelbench_valid_level1}
  \centering
    \resizebox{0.98\textwidth}{!}{
        \begin{tabular}{lcccccc}
        \toprule
        \multirow{2}{*}{\textbf{Model}} 
        & \multirow{2}{*}{\textbf{\#Params}} 
        & \multicolumn{4}{c}{\textbf{\textsc{Level1 (robust verifier)}}} 
        & \multicolumn{1}{c}{\textbf{\textsc{Level1 (w/o robust verifier)}}} \\
        \cmidrule(lr){3-6}
        \cmidrule(lr){7-7}
        & & \textbf{valid} & \textbf{compiled}~/~\textbf{correct} 
        & $\textbf{fast}_1$~/~$\textbf{fast}_2$ & \textbf{mean speedup} 
        & \textbf{compiled}~/~\textbf{correct} 
        \\         
        \midrule
        Qwen3 (base)  &   $8$B    & $99.0$ &$99.0$~/~$23.0$ & $9.0$~/~$5.0$ & $0.41$ & $100.0$~/~$25.0$ \\
        Qwen3       &  $14$B  & $100.0$ &$99.0$~/~$35.0$ & $9.0$~/~$7.0$ & $0.48$ & $100.0$~/~$35.0$  \\
        Qwen3       &  $32$B  & $100.0$ & $100.0$~/~$62.0$  & $39.0$~/~$16.0$ & $0.93$ & $100.0$~/~$67.0$   \\ \midrule 
        KernelLLM & $8$B &  $42.0$  & $40.0$~/~$20.0$ & $0.0$~/~$0.0$ & $0.05$ & $99.0$~/~$34.0$  \\
        AutoTriton & $8$B &$97.0$  & $92.0$~/~$57.0$ & $25.0$~/~$10.0$ & $0.95$ & $100.0$~/~$87.0$  \\
        \ours(D)   & $8$B  &   $100.0$    & $\mathbf{100.0}$~/~$78.0$     &  $36.0$~/~$13.0$ & $1.10$ & $100.0$~/~$81.0$ \\
        \ours(D) + IA   & $8$B  &   $100.0$    & $99.0$~/~$78.0$     &  $34.0$~/~$15.0$ & $1.05$ & $100.0$~/~$81.0$ \\        
        \ \ \  SFT w. DeepSeek-R1 (w.o. RL)  & $8$B & $100.0$ & $\mathbf{100.0}$~/~$54.0$   & $28.0$~/~$12.0$ & $0.88$ & $100.0$~/~$60.0$  \\ 
        \ours (G)    & $8$B  &   $100.0$    & $\mathbf{100.0}$~/~$76.0$ &   $35.0$~/~$17.0$ & $1.09$ & $100.0$~/~$78.0$\\
        \ours (G) + IA  & $8$B  &   $100.0$    & $\mathbf{100.0}$~/~$\mathbf{88.0}$     &  $\mathbf{41.0}$~/~$\mathbf{22.0}$ & $\mathbf{1.26}$ & $100.0$~/~$88.0$  \\
        \ \ \ SFT w. GPT-OSS 120B (w.o. RL)   & $8$B & $100.0$    & $\mathbf{100.0}$~/~$64.0$ & $26.0$~/~$18.0$ & $1.03$ & $100.0$~/~$73.0$\\ 
        \midrule\midrule
        Claude-3.7  & -  & $100.0$ & $100.0$~/~$63.0$ & $29.0$~/~$16.0$ & $1.06$ & $100.0$~/~$73.0$ \\
        DeepSeek-R1  & $685$B  & $100.0$ & $100.0$~/~$69.0$ & $28.0$~/~$13.0$ & $1.04$ & $100.0$~/~$74.0$ \\
        GPT-OSS  & $120$B  & $100.0$ & $100.0$~/~$81.0$ & $33.0$~/~$22.0$ & $1.20$ & $100.0$~/~$87.0$\\
        \bottomrule
        \toprule
        \multirow{2}{*}{\textbf{Model}} 
        & \multirow{2}{*}{\textbf{\#Params}} 
        & \multicolumn{4}{c}{\textbf{\textsc{Level2 (robust verifier)}}} 
        & \multicolumn{1}{c}{\textbf{\textsc{Level2 (w/o robust verifier)}}} \\
        \cmidrule(lr){3-6}
        \cmidrule(lr){7-7}
        & & \textbf{valid} & \textbf{compiled}~/~\textbf{correct} 
        & $\textbf{fast}_1$~/~$\textbf{fast}_2$ & \textbf{mean speedup} 
        & \textbf{compiled}~/~\textbf{correct} 
        \\         
        \midrule
          Qwen3 (base) &   $8$B    &  $38.0$ & $37.0$~/~$0.0$ & $0.0$~/~$0.0$ & $0.00$ & $100.0$~/~$56.0$ \\
          Qwen3       &  $14$B  & $40.0$ & $38.0$~/~$1.0$ & $0.0$~/~$0.0$ & $0.00$ & $100.0$~/~$77.0$ \\
          Qwen3       &  $32$B  & $36.0$ & $35.0$~/~$2.0$ & $2.0$~/~$1.0$  & $0.10$ & $100.0$~/~$68.0$ \\
          \midrule
          KernelLLM & $8$B &  $3.0$ & $0.0$~/~$0.0$ & $0.0$~/~$0.0$ & $0.00$ & $100.0$~/~$4.0$  \\
          AutoTriton & $8$B & $11.0$ & $10.0$~/~$1.0$ & $0.0$~/~$0.0$ & $0.00$ & $100.0$~/~$94.0$ \\
          \ours (D)  & $8$B  &  $63.0$ & $63.0$~/~$23.0$   &  $12.0$~/~$3.0$  & $0.32$ & $100.0$~/~$77.0$ \\
          \ours (D) + IA  & $8$B  &   $52.0$    & $51.0$~/~$24.0$     &  $16.0$~/~$7.0$ & $0.31$ & $100.0$~/~$79.0$ \\        
          \ \ \  SFT w. DeepSeek-R1 (w.o. RL)  & $8$B & $55.0$ & $51.0$~/~$13.0$   & $10.0$~/~$3.0$ & $0.31$ & $100.0$~/~$67.0$ \\ 
          \ours (G)    & $8$B  &  $91.0$ &  $91.0$~/~$23.0$  &  $15.0$~/~$7.0$ & $0.29$  & $98.0$~/~$30.0$\\
          \ours (G) + IA & $8$B  &  $\mathbf{93.0}$ & $\mathbf{93.0}$~/~$\mathbf{28.0}$   &  $\mathbf{19.0}$~/~$\mathbf{10.0}$ & $\mathbf{0.48}$ & $96.0$~/~$37.0$ \\
          \ \ \ SFT w. GPT-OSS 120B (w.o. RL)   & $8$B & $88.0$ & $88.0$~/~$13.0$ & $9.0$~/~$3.0$ & $0.21$ & $97.0$~/~$25.0$ \\ 
          \midrule \midrule 
          Claude-3.7  & -  & $34.0$ & $34.0$~/~$14.0$ & $4.0$~/~$1.0$ & $0.10$ & $100.0$~/~$65.0$ \\
          DeepSeek-R1  & 685B  & $31.0$ & $31.0$~/~$14.0$ & $9.0$~/~$4.0$& $0.24$ & $100.0$~/~$79.0$ \\
          GPT-OSS  & 120B  & $39.0$ & $39.0$~/~$15.0$ & $12.0$~/~$4.0$ & $0.25$ & $100.0$~/~$81.0$\\
          \bottomrule
    \end{tabular}
    }
    \vspace{-3mm}
\end{table*}

\section{Experiments}
\label{sec:exp}
This section provides the detailed recipe of training and evaluation of \ours, followed by the main results and ablation studies.
\subsection{Training and Evaluation Setups}
\label{sec:training_setups}

\paragraph{Training Configuration.}
We implement the training pipeline using the VeRL framework \citep{sheng2025hybridflow} and initialize all experiments from Qwen3-8B \citep{qwen3tech}. Unless noted otherwise, RL training starts from the SFT checkpoint distilled from DeepSeek-R1 (denoted by  \ours (D)) and uses the reward $r$ in \eqref{eq:reward} with $\alpha^*=0.1$. 
SFT is performed with batch size $16$ and learning rate $1\times10^{-5}$; RL uses batch size $32$ and learning rate $1\times10^{-6}$. 
We denote the model fine-tuned from the GPT-OSS $120$B SFT checkpoint as \ours (G); most training hyperparameters were kept the same (see Appendix~\ref{sec:appendix:hps} for details).

\paragraph{Evaluation Benchmarks.}
We evaluate \ours on KernelBench \citep{li2025tritonbench}\footnote{We use the Triton backend version of KernelBench from \url{https://github.com/ScalingIntelligence/KernelBench/pull/35}.}. KernelBench offers an evaluation framework covering 250 tasks, divided into Level 1 (100 single-kernel tasks, such as convolution), Level 2 (100 simple fusion tasks, such as conv+bias+ReLU), and Level 3 (50 full architecture tasks, such as MobileNet), to assess LLM proficiency in generating efficient CUDA kernels. We conduct experiments mainly on the Level 1 and Level 2 tasks from KernelBench. The prompts used for these benchmarks are provided in Appendix~\ref{sec:kernelbench_example}.

\paragraph{Metrics.}
We evaluate the performance of LLMs for generating Triton code in terms of (1) Validity (syntax and functionality); (2) Correctness (compilation and correct output); (3) Speedup (relative execution time improvement).  
We report \texttt{fast}$_1$ and \texttt{fast}$_2$ to indicate the model's ability to generate Triton code that is at least as fast as or twice as fast as the reference PyTorch implementation, respectively. The formal definition of metrics is provided in Appendix~\ref{sec:metrics}. We measure the pass@$k$ metrics for each aspect, which indicates the ratio of generating at least one successful solution among $k$ sampled attempts. We use $k=10$ as a default unless specified. We test both Triton codes and reference PyTorch codes on an NVIDIA L40S.

\paragraph{Baselines.}
We compare \ours with several baselines, including KernelLLM \citep{kernelllm2025} and AutoTriton \citep{li2025autotriton}, which are fine-tuned LLMs specifically for Triton programming. We also include our base model Qwen3-8B \citep{qwen3tech} without any fine-tuning, fine-tuned Qwen3-8B only after SFT, and larger Qwen3 models (e.g., Qwen3-14B and Qwen3-32B). Additionally,
we evaluate Claude-3.7 \citep{claude37} with unknown model size and large model classes beyond 100B (e.g., GPT-OSS 120B \citep{openai2025gptoss120b}, DeepSeek-R1-0528 \citep{guo2025deepseek}) as a reference.

\subsection{Main Experiment Results}
\label{sec:main_results}
\paragraph{Correctness and Speedup Evaluation on KernelBench.}
The left side of Table~\ref{tab:kernelbench_valid_level1} shows the pass@10 results using the robust verifier (syntax and functionality) on KernelBench Level 1 (single kernel) and Level 2 (fusion) tasks. \ours consistently outperforms other baselines, including Triton-specific models (KernelLLM and AutoTriton) with fewer than 32B parameters, in terms of validity, correctness, and speedup on both task levels.
In particular, \ours outperforms AutoTriton, which also employs RL, by delivering higher correctness and greater speedup, highlighting the effectiveness of our RL framework with its robust verifiers and hierarchical reward decomposition.
Importantly, RL provides significant gains beyond knowledge distillation from large models (DeepSeek-R1 and GPT-OSS $120$B), boosting correctness by over $20\%$ and increasing \texttt{fast}$_1$ by roughly $10$--$15\%$.  
Furthermore, \ours achieves on-par or better performance even compared to frontier models, highlighting the effectiveness of our approach in enabling smaller models to excel in specialized code generation tasks. 

In Appendix~\ref{sec:passk_results}, we present additional evaluation results across varying $k$ values for pass@$k$ metrics. These results demonstrate that \ours consistently outperforms baselines as $k$ increases, indicating that our model generates more diverse candidate solutions and effectively leverages additional sampling during inference.

\begin{table}[ht!] \small
  \caption{
Comparison of hierarchical versus uniform reward designs on KernelBench Level 1 tasks. All metrics are reported as pass@10 (\%).
Reward type specifies whether the single reward is assigned uniformly across all tokens (Uniform) as in \eqref{eq:uniform_reward_mixture}, or the reward signal is decomposed by token class (Hierarchical) as in \eqref{eq:reward}. 
The hyperparameter $\beta$ controls the proportion of correctness in the single reward function applied to all tokens. The models are trained from the same SFT checkpoint distilled from DeepSeek-R1.
  }
  \label{tab:reward_shape_ablation}
  \centering
  \resizebox{0.98\columnwidth}{!}{
    \begin{tabular}{llccc}
    \toprule
    \textbf{Reward Type} & $\bm{\beta}$
    & \textbf{valid} & \textbf{compiled}~/~\textbf{correct} & $\textbf{fast}_\textbf{1}$~/~$\textbf{fast}_\textbf{2}$ \\
    \midrule
    Hierarchical ($\alpha^*$)& - &    $100.0$    & $100.0$~/~$\mathbf{78.0}$     &  $\mathbf{36.0}$~/~$13.0$   \\ \midrule
    Uniform & $0.0$ & $100.0$  & $100.0$~/~$65.0$ & $33.0$~/~$\mathbf{15.0}$ \\
    Uniform  & $0.3$ & $100.0$  & $100.0$~/~$68.0$ & $29.0$~/~$14.0$ \\
    Uniform  & $0.5$ & $100.0$  & $100.0$~/~$68.0$ & $35.0$~/~$13.0$ \\
    Uniform  & $0.7$ &  $100.0$  & $100.0$~/~$70.0$ & $34.0$~/~$12.0$ \\  
    Uniform  & $0.9$ & $100.0$  & $100.0$~/~$75.0$ & $31.0$~/~$11.0$ \\  
    Uniform  & $1.0$ & $100.0$  & $100.0$~/~$70.0$ & $35.0$~/~$12.0$ \\
    \bottomrule
  \end{tabular}
  }
  \vspace{-3mm}
\end{table}

\begin{table*}[ht!] \small
 \caption{Ablation study on data mixture for RL training of \ours, where the performance is evaluated on KernelBench level 1 and level 2 tasks. The models are trained from the same SFT checkpoint distilled from DeepSeek-R1.
  }
  \label{tab:data_mixture}
  \centering
  \resizebox{0.85\textwidth}{!}{
    \begin{tabular}{lccccccc}
    \toprule
    \multirow{2}{*}{\textbf{Train Data}} & \multirow{2}{*}{\textbf{Mixing Prob.}}
    & \multicolumn{3}{c}{\textbf{\textsc{Level1}}} 
    & \multicolumn{3}{c}{\textbf{\textsc{Level2}}} \\
    \cmidrule(lr){3-6}
    \cmidrule(lr){6-8}
     \textbf{Mixture} & $\bm{p}$ & \textbf{valid} & \textbf{compiled}~/~\textbf{correct} & $\textbf{fast}_\textbf{1}$~/~$\textbf{fast}_\textbf{2}$ 
      & \textbf{valid} & \textbf{compiled}~/~\textbf{correct} & $\textbf{fast}_\textbf{1}$~/~$\textbf{fast}_\textbf{2}$ 
    \\         
    \midrule
    Level 1 & $[1, 0]$ &  $100.0$    & $\mathbf{100.0}$~/~$\mathbf{78.0}$     &  $\mathbf{36.0}$~/~$13.0$ & $\mathbf{63.0}$ & $\mathbf{63.0}$~/~$\mathbf{23.0}$   &  $\mathbf{12.0}$~/~$3.0$ \\
    Level 1+2 & $[0.5, 0.5]$ &$100.0$ & $99.0$~/~$73.0$ & $33.0$~/~$\mathbf{14.0}$ & $\mathbf{63.0}$ & $\mathbf{63.0}$~/~$22.0$ & $9.0$~/~$3.0$ \\
    Level 2 & $[0, 1]$ &$100.0$ & $\mathbf{100.0}$~/~$70.0$ & $30.0$~/~$10.0$ & $50.0$ & $50.0$~/~$16.0$ & $9.0$~/~$\mathbf{4.0}$ \\
    \bottomrule
  \end{tabular}
  }
  \vspace{-3mm}
\end{table*}

\paragraph{Robust Verification Prevents Reward Hacking.}
In Figure~\ref{fig:reward_hacking}, we analyze the validity of Triton codes generated by fine-tuned models to understand the types of errors each model is prone to. 
Although \ours and other fine-tuned baselines (KernelLLM and AutoTriton) learn to generate more valid codes after fine-tuning, a more detailed breakdown reveals that they exhibit a much higher proportion of functionally invalid codes (red). 
In contrast, \ours generates significantly fewer invalid codes in terms of both syntax and functionality after fine-tuning, demonstrating the effectiveness of our robust verification in enhancing code quality without reward hacking.

Moreover, Table~\ref{tab:kernelbench_valid_level1} highlights how heavily prior models relied on cheating shortcuts. 
Without functionality verification (i.e., w/o the robust verifier), AutoTriton's correctness jumps from 57\% to 87\%, revealing its tendency to exploit loopholes by superficially meeting syntax checks rather than producing truly functional Triton codes.
In contrast, \ours shows only a slight increase ($\le 3$\%), suggesting it learns to generate genuine code without relying on shortcuts.

\paragraph{Effectiveness of HRD: Hierarchical vs. Uniform.}
To validate the effectiveness of our hierarchical reward decomposition, we compare it with a \emph{uniform} reward assignment approach, where a single reward is uniformly applied to all tokens without distinguishing between plan and code tokens, which is the reward design chosen by \citet{li2025autotriton,baronio2025multi}. 
To generalize the reward function used in \citet{li2025autotriton} and \citet{baronio2025multi}, we define a uniform reward as a weighted combination of correctness and speedup conditioned on validity,
which can be expressed as 
 \begin{equation} \label{eq:uniform_reward_mixture}
 \scalebox{0.88}{ 
   $\displaystyle 
   \begin{aligned}
 r_{i,j} &= \beta \cdot R^{\text{correct}}(g_i, o_{i,j})+ (1-\beta) \cdot R^{\text{speedup}}(g_i, o_{i,j}),
  \end{aligned}
   $
 }
 \end{equation}
where $\beta \in [0,1]$ is a hyperparameter that controls the ratio of correctness in the reward mixture.
In \citet{li2025autotriton}, the reward is computed only based on correctness, which corresponds to $\beta=1.0$, while in \citet{baronio2025multi}, the reward is defined as the sum of correctness and speedup with some fixed weights.
We defer the detailed GRPO formulation for uniform reward to Appendix~\ref{sec:appendix:uniform_reward}.
We compare \ours using hierarchical reward decomposition against models trained with uniform reward designs for various $\beta$ values, keeping all other RL training settings (GRPO algorithm, hyperparameters, and pre-RL fine-tuning) consistent.

Table~\ref{tab:reward_shape_ablation} demonstrates that hierarchical reward decomposition consistently yields higher correctness and speedup than any uniform reward configuration. 
While prior works \citep{li2025autotriton,baronio2025multi} explored single reward functions with various mixtures of correctness and speedup, our results demonstrate that such uniform designs have limitations in achieving the best model performance. 
This suggests that decoupling rewards by token class provides more targeted learning signals, better aligning with the distinct roles of planning and coding and leading to higher-quality Triton code generation.

\paragraph{Ablation Study on Plan-to-Code Update Ratio ($\alpha$).}
We also analyze the effect of the hyperparameter $\alpha$, which controls the relative update rate of planning versus coding actions during RL training. Smaller $\alpha$ values slow down plan token updates, allowing coding actions to adapt to more stable planning distributions. As shown in Table~\ref{tab:alpha_ablation}, $\alpha=0.1$ achieves the best overall performance, while higher values (e.g., $\alpha=1.0$) lead to instability and reduce correctness and speedup. Conversely, setting $\alpha=0.0$, disabling plan updates, also degrades results, indicating that some plan adaptation is necessary. These findings highlight the importance of balancing plan and code updates to avoid premature convergence and maximize Triton code quality.

\begin{table}[ht!] \small
\caption{Ablation study of the hyperparameter $\alpha$ in \ours on KernelBench Level 1 tasks. All metrics are reported as pass@10 (\%). 
$\alpha$ is the weighting factor, determining how quickly plan tokens are updated relative to code tokens during training.
The default configuration uses $\alpha^* = 0.1$ and the models are trained from the same SFT checkpoint distilled from DeepSeek-R1.}
  \label{tab:alpha_ablation}
  \centering
  \resizebox{0.9\columnwidth}{!}{
    \begin{tabular}{lcccc}
    \toprule
    \multirow{1}{*}{$\alpha$} 
    & \textbf{valid} & \textbf{compiled}~/~\textbf{correct} & $\textbf{fast}_\textbf{1}$~/~$\textbf{fast}_\textbf{2}$ \\
    \midrule
    $0.0$ &  $100.0$ & $100.0$~/$72.0$ & $32.0$~/~$12.0$ \\
    $0.1$ ($\alpha^{*}$) &  $100.0$    & $100.0$~/~$\mathbf{78.0}$     &  $\mathbf{36.0}$~/~$13.0$    \\
    $0.5$& $100.0$  & $100.0$~/~$75.0$ & $35.0$~/~$12.0$ \\
    $0.8$ & $100.0$  & $99.0$~/~$70.0$ & $33.0$~/~$10.0$ \\
    $1.0$ & $100.0$  & $100.0$~/~$71.0$ & $31.0$~/~$11.0$ \\
    \bottomrule
  \end{tabular}
  }
  \vspace{-3mm}
\end{table}

\paragraph{Comparison of Difficulty-Based Data Mixtures.}
We explore different data mixtures for RL training of \ours to understand how training data composition affects model performance. Our training dataset consists of two levels of tasks, with the ratio controlled by a data mixture probability vector $\bm{p} = [p_1, p_2]$, where $p_1$ and $p_2$ represent the probabilities of sampling Level 1 and Level 2 tasks, respectively. 
We evaluate \ours on KernelBench Level 1 and 2 tasks using three data mixing strategies for RL training: training exclusively on Level 1 tasks ($\bm{p}=[1, 0]$), exclusively on Level 2 tasks ($\bm{p}=[0, 1]$), and a balanced mixture of both ($\bm{p}=[0.5, 0.5]$), as summarized in Table~\ref{tab:data_mixture}.
Interestingly, training exclusively on Level 1 tasks ($\bm{p}=[1, 0]$) yields the best correctness and $\text{fast}_1$ on both Level 1 and Level 2 evaluations. Including Level 2 tasks during RL training provides only marginal improvements or worsens performance on Level 2 tasks. This may be because Level 2 tasks are inherently more complex, and reward signals from those tasks are very sparse, making it difficult for the model to learn effectively. Consequently, it suggests that focusing on fundamental single-operation tasks builds more robust Triton block generation skills that transfer effectively to both simple and complex fusion scenarios, rather than directly training on the more challenging fusion problems themselves.

\paragraph{Diverse Input Augmentation Improves Triton Quality.}
To assess the impact of input augmentation (IA) used for RL data construction (Section~\ref{sec:rl_data_construction}), we compare \ours trained with and without IA under identical settings in Table~\ref{tab:kernelbench_valid_level1}. Especially, \ours (G) + IA substantially improves over \ours (G) without IA on both Level 1 and Level 2 tasks, correctness increases by $12\%$ and \texttt{fast}$_1$ by $6\%$ on KernelBench Level 1. This demonstrates that IA diversifies the training data and helps the model generalize to produce higher-quality Triton kernel codes. 
To demonstrate robustness to diverse input shapes, we evaluate \ours with IA on augmented KernelBench tasks with varied input shapes (see Appendix~\ref{sec:input_augmentation_ablation}).

\paragraph{Challenges and Future Directions.}
Table~\ref{tab:kernelbench_valid_level1} demonstrates that \ours maintains its advantage over both much larger ($>$100B) models and 8B-scale Triton-specific models on KernelBench Level 2 fusion tasks. However, consistent with prior work \cite{li2025autotriton,kernelllm2025}, fusion task generation remains a significant open challenge for all models, with relatively lower performance compared to Level 1 single-kernel tasks. Our analysis reveals two primary factors: (1) models frequently produce partial implementations that delegate computation to high-level PyTorch APIs, and (2) the sparsity of successful fusion implementations leads to weak reward signals during training (Table~\ref{tab:data_mixture}).
Despite these challenges, \ours achieves the best results among models of comparable or larger scale, and our RL framework, with its robust verification and hierarchical reward mechanisms, provides essential foundations for future work in complex kernel generation.
As kernel patterns grow more diverse, extending the verifier with formal guarantees, evaluating across additional benchmarks and accelerator targets, and exploring how HRD interacts with larger model scales represent natural next steps toward a more general hardware-aware RL paradigm.

\section{Conclusion}
In this work, we introduce \ours, a specialized LLM for Triton code generation, trained with a novel RL framework featuring robust verifiable rewards and hierarchical reward assignment. Our experiments on KernelBench show that \ours surpasses existing fine-tuned Triton models in validity, correctness, and efficiency. Ablation studies demonstrate that both robust reward design and hierarchical reward assignment are essential for achieving correctness and efficiency.

\section*{Impact Statement}

This paper presents work whose goal is to advance the field of Machine
Learning. There are many potential societal consequences of our work, none of which we feel must be specifically highlighted here.

\bibliography{reference}
\bibliographystyle{icml2026}

\newpage
\appendix
\onecolumn

\section{Related Work}
\label{appendix:related_work}

\subsection{LLM for Kernel Generation}
The exponential growth in demand for GPU computing resources has driven the need for highly optimized GPU kernels that improves computational efficiency. However, writing efficient GPU kernels is a complex and time-consuming task that requires specialized knowledge of GPU architectures and programming models. This has spurred significant interest in leveraging Large Language Models (LLMs), for automated kernel generation, especially for CUDA and Triton \citep{shao2024deepseekmath,ouyang2025kernelbench,li2025tritonbench,nvidia2025deepseek_kernelgen}. While these general-purpose models excel at a variety of programming tasks, they often struggle with custom kernel generation, achieving low success rates on specialized gpu programming tasks \citep{ouyang2025kernelbench}, highlighting the need for domain-specific models tailored to kernel synthesis.

For CUDA kernel generation, \citet{ouyang2025kernelbench} introduced \textsc{KernelBench}, an open-source framework
for evaluating LMs’ ability to write fast and correct kernels on a suite of 250 carefully selected PyTorch
ML workloads. Furthermore, \citet{lange2025ai} presented an agentic framework, which leverages LLMs to translate PyTorch code into CUDA kernels and iteratively optimize them using performance feedback. Additionally, several works have focused on fine-tuning LLMs tailored for CUDA kernel generation. For example, Kevin-32B \citep{baronio2025multi} is a 32B parameter model fine-tuned via multi-turn RL to enhance kernel generation through self-refinement, and CUDA-L1 \citep{li2025cudal1} applies contrastive reinforcement learning to DeepSeek-V3-671B, achieving notable speedup improvements in CUDA optimization tasks.

Another line of research focuses on Triton kernel generation. \citet{li2025tritonbench} introduced \textsc{TritonBench}, providing evaluations of LLMs on Triton programming tasks and highlighting the challenges of Triton's domain-specific language and GPU programming complexity.
To further enhance LLMs' capabilities in Triton programming, \citet{kernelllm2025} have introduced KernelLLM, a fine-tuned model of Llama3.1-8B-Instruct via supervised fine-tuning with Pytorch and Triton code pairs in KernelBook \citet{kernelbook2025}, but its performance is limited by the quality of training data.
Similarly, \citet{li2025autotriton} introduced AutoTriton, a model fine-tuned specifically for Triton programming from Seed-Coder-8B-Reasoning \citet{zhang2025seed}, which achieves improved performance via SFT and RL with verifiable rewards based on correctness and rule-based Triton syntax verification, which may have limited improvement in runtime efficiency due to correctness-focused rewards.
Both KernelLLM and AutoTriton are concurrent works developed alongside our work, and we provide a detailed comparison in Section~\ref{sec:main_results}.

\subsection{Reinforcement Learning with Verifiable Rewards}
Reinforcement Learning (RL) has become a key technique for training Large Language Models (LLMs), especially in domains where verifiable reward signals are available. Unlike supervised fine-tuning (SFT), which relies on curated examples, RL enables models to learn through trial and error, guided solely by reward feedback. This makes the design of accurate reward functions critical, as the model's behavior is shaped entirely by the reward signal. As a result, RL with verifiable rewards (RLVR) \citep{lambert2025tulu3,kimiteam2025kimik15,guo2025deepseek} has gained significant traction in applications like mathematics and code generation \citep{shao2024deepseekmath,AlphaCode2022}, where external verification is feasible through solution correctness or unit test outcomes.

In math and coding applications, the reward can be directly computed solely based on the final outcomes when ground-truth answers or unit tests are available. For tasks where validation is not available or noisy, rule-based verification or LLM-based judges can be employed to verify the quality of generated content \citep{guha2025openthoughts,guo2025deepseek}.
For coding tasks, unit tests are commonly used to measure whether generated code meets the specified requirements \citep{CodeRL2022,ouyang2025kernelbench}. However, unit tests often fail to cover edge cases or fully capture the problem requirements, leading to potential ``reward hacking'' \citep{skalse2022defining} where the model generates code that passes the tests but does not genuinely solve the task \citep{sharma2024critical,gao2024designing}. Such reward hacking has been observed in kernel generation tasks, where models produce superficially correct codes passing unit tests by using high-level Pytorch modules instead of implementing custom kernels. To address this, some works \citet{li2025autotriton,baronio2025multi} have introduced rule-based verification, which checks kernel syntax or use of specific high-level modules.

\section{Notations}
\label{sec:appendix}

The following notations will be used throughout this paper. For notational simplicity, we denote any function $f(q,o_i)$ as $f_i$ when the context is clear.
\begin{itemize}
  \item $q$: prompt given to the model, defining a task to implement in Triton
  \item $o$: output sequence generated by the model, which includes both reasoning trace and Triton code
  \item $\pi_\theta$: policy model with parameters $\theta$
  \item $G$: group size for GRPO
  \item $o_{i,j}$: $j$-th sample in the group $G$ for the $i$-th prompt, which includes a reasoning trace that provides the "plan" for Triton code optimization and implementation and the final "Triton code", i.e. $o_{i,j} = \{o_{i, j}^{\text{plan}}, o_{i,j}^{\text{code}}\}$
  \item $T_{i,j}^c$: set of token indices corresponding to token class $c \in \{\text{plan}, \text{code}\}$ in the $j$-th sample of the $i$-th prompt
  \item $r_{i,j}^{c}$: reward function for token class $c \in \{\text{plan}, \text{code}\}$.
  \item $A_{i,j}^c = r_{i,j}^c - \frac{1}{G} \sum_{j=1}^G r_{i,j}^c$: token-level advantage of the $j$-th sample for prompt $q_i$ belonging to token class $c \in \{\text{plan}, \text{code}\}$.
\end{itemize}

\section{Metrics}
\label{sec:metrics}
We provide the formal definitions of the evaluation metrics used in this paper. Given a set of $N$ tasks $\{g_n\}_{n=1}^N$ and $k$ samples $\{o_{n,i}\}_{i=1}^k$ generated by the model for each task, we define the following metrics:
\begin{equation}
\scalebox{0.9}{
$\displaystyle
\begin{aligned}
&\text{valid} = \frac{1}{N} \sum_{n=1}^{N} \max_{i \in [k]}\mathbbm{1}(\texttt{syntax}(g_n, o_{n,i}) \cdot \texttt{func}(g_n, o_{n,i}) = 1) \\
&\text{compiled} = \frac{1}{N} \sum_{n=1}^{N} \max_{i \in [k]}\mathbbm{1}(\texttt{syntax}(g_n, o_{n,i}) \cdot \texttt{func}(g_n, o_{n,i}) \cdot \texttt{compiled}(g_n, o_{n,i}) = 1) \\
&\text{correct} = \frac{1}{N} \sum_{n=1}^{N} \max_{i \in [k]}\mathbbm{1}(\texttt{syntax}(g_n, o_{n,i}) \cdot \texttt{func}(g_n, o_{n,i}) \cdot \texttt{correct}(g_n, o_{n,i}) = 1) \\
&\text{fast}_p = \frac{1}{N} \sum_{n=1}^{N} \max_{i \in [k]}\mathbbm{1}(\texttt{syntax}(g_n, o_{n,i}) \cdot \texttt{func}(g_n, o_{n,i}) \cdot \texttt{correct}(g_n, o_{n,i}) \cdot \texttt{speedup}(g_n, o_{n,i}) > p) \\
&\text{mean\_speedup} = \frac{1}{N} \sum_{n=1}^{N} \max_{i \in [k]}  (\texttt{syntax}(g_n, o_{n,i}) \cdot \texttt{func}(g_n, o_{n,i}) \cdot \texttt{correct}(g_n, o_{n,i}) \cdot \texttt{speedup}(g_n, o_{n,i}) ) \\
\end{aligned}
$
}
\end{equation} 
To measure correctness, we execute the generated Triton code for five randomly generated inputs and compare the outputs against those from the reference PyTorch code. If all outputs match, we consider the generated code correct. 
To measure speedup, we executes a reference code and generated Triton code on the same hardware setup (NVIDIA L40S) for 10 times for a randomly generated input and compute the ratio of their mean execution times.

\section{GRPO with uniform reward assignment}
\label{sec:appendix:uniform_reward}
Here we provide the detailed formulation of GRPO with uniform reward assignment, which is used in our experiments (Section~\ref{sec:main_results}) to compare with our proposed hierarchical reward assignment. The GRPO objective with uniform reward assignment is defined as
\begin{equation}
\scalebox{1.0}{ 
  $\displaystyle 
  \begin{aligned}
  \mathcal{J}_{\text{\tiny GRPO}}(\theta) &= \mathbb{E}_{q_i \sim \mathcal{D}_{RL}, ~ \{o_{i,j} \}_{j=1}^G \sim \pi_{\theta_{old}}(\cdot|q_i) } \left [ \mathcal{L}_{\text{\tiny GRPO}} (\theta, i) \right ]
  \end{aligned}
  $
}
\end{equation}
where $\pi_\theta$ and $\pi_{\theta_{old}}$ are the policy model and reference model, $q_i$ denotes a prompt given to the model, defining a task to implement in Triton, and $o_{i,j}$ represents $i$-th response generated by the model for $q_i$ in the group $G$. The GRPO losses $\mathcal{L} (\theta,i)$ is computed as
\begin{equation}
\scalebox{0.8}{ 
$\displaystyle 
    \begin{aligned}
    \mathcal{L}_{\text{\tiny GRPO}} (\theta,i) &= \frac{1}{G} \sum_{j=1}^G \frac{1}{|o_{i,j}|} \sum_{t=1}^{|o_{i,j}|} A_{i,j} \cdot \min \left\{   \frac{\pi_\theta(o_{i,j,t}|q, o_{i,j,<t})}{\pi_{\theta_{old}}(o_{i,j,t}|q, o_{i,j,<t})} , \text{clip} \left( \frac{\pi_\theta(o_{i,j,t}|q, o_{i,j,<t})}{\pi_{\theta_{old}}(o_{i,j,t}|q, o_{i,j,<t})}, 1-\epsilon, 1+\epsilon \right)  \right\} ,
    \end{aligned}
    $
}
\end{equation}
where $A_{i,j}$ denotes the group-wise advantage uniformly applied to all tokens, computed as $A_{i,j} = r_{i,j} - \frac{1}{G} \sum_{j=1}^G r_{i,j}$, with $r_{i,j}$ being the reward for the $i$-th sample. In the uniform reward assignment, we define the reward as a weighted combination of correctness and speedup conditioned on validity, which can be expressed as
\begin{equation} 
 \scalebox{1.0}{ 
   $\displaystyle 
   \begin{aligned}
 r_{i,j} &= \beta \cdot R^{\text{correct}}(g_i, o_{i,j})+ (1-\beta) \cdot R^{\text{speedup}}(g_i, o_{i,j}),
  \end{aligned}
   $
 }
 \end{equation}
where $\beta \in [0,1]$ is a hyperparameter that the ratio of correctness in the reward mixture.

\section{Training Hyperparameters and Details}
\label{sec:appendix:hps}
For generating reasoning and code traces for SFT, we use temperature=0.6, top\_p=0.95 for DeepSeek-R1 \citep{guo2025deepseek} and temperature=1.0, top\_p=1.0 for GPT-OSS 120B \citep{openai2025gptoss120b}. 
To label difficulty, we further label tasks into three difficulty levels using Qwen3-235B-Instruct \citep{qwen3tech} (temperature=0.7, top\_p=0.8).

For SFT, we train for 2 epochs on DeepSeek-R1-generated data with a maximum sequence length of $12,288$ tokens and 3 epochs on GPT-OSS 120B-generated data with a maximum sequence length of $16,384$ tokens, all with a batch size of 16 and a learning rate of $1 \times 10^{-5}$.

For RL, we train the model for 2 epochs using the VeRL framework  \citep{sheng2025hybridflow} with a batch size of 32, a learning rate of $1 \times 10^{-6}$, a clip ratio of $\epsilon = 0.2$ for the GRPO objectives, the maximum prompt length $2,048$, and the maximum response length $16,384$. We use 8 NVIDIA A100 80GB GPUs for both SFT and RL training. 

Note that \ours (G), a model trained with SFT checkpoint distilled from GPT-OSS $120$B, was trained with 8 NVIDIA H200 GPUs due to resource constraints. Also, we use KernelBook \citep{kernelbook2025} tasks where input generating functions are augmented using GPT-OSS $120$B for RL training. 

\section{Additional Experiment Results}

\subsection{Pass@k Results}
\label{sec:passk_results}

\begin{figure}[h]
  \centering
  \includegraphics[width=0.47\linewidth]{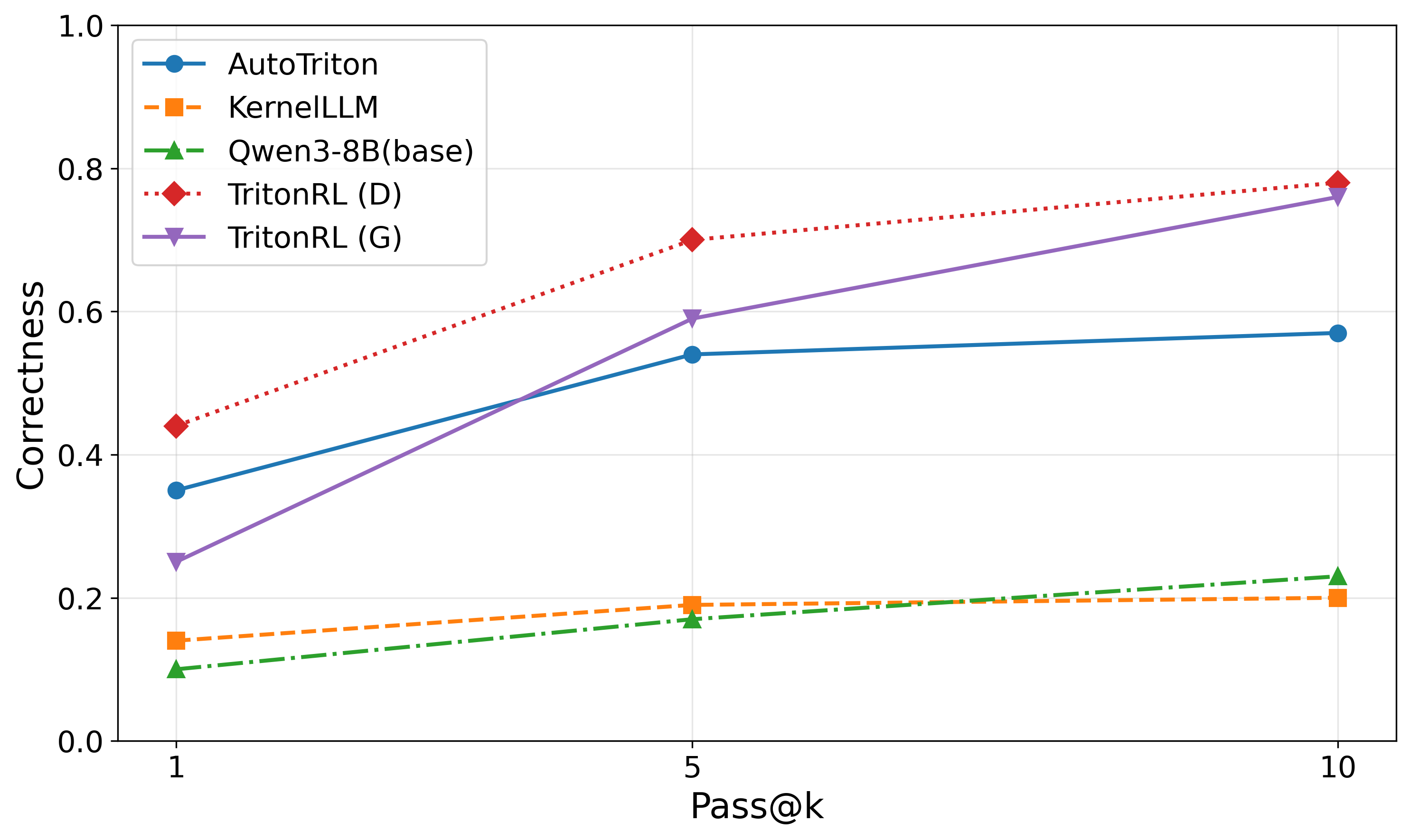}
  \hspace{0.1cm}
  \includegraphics[width=0.47\linewidth]{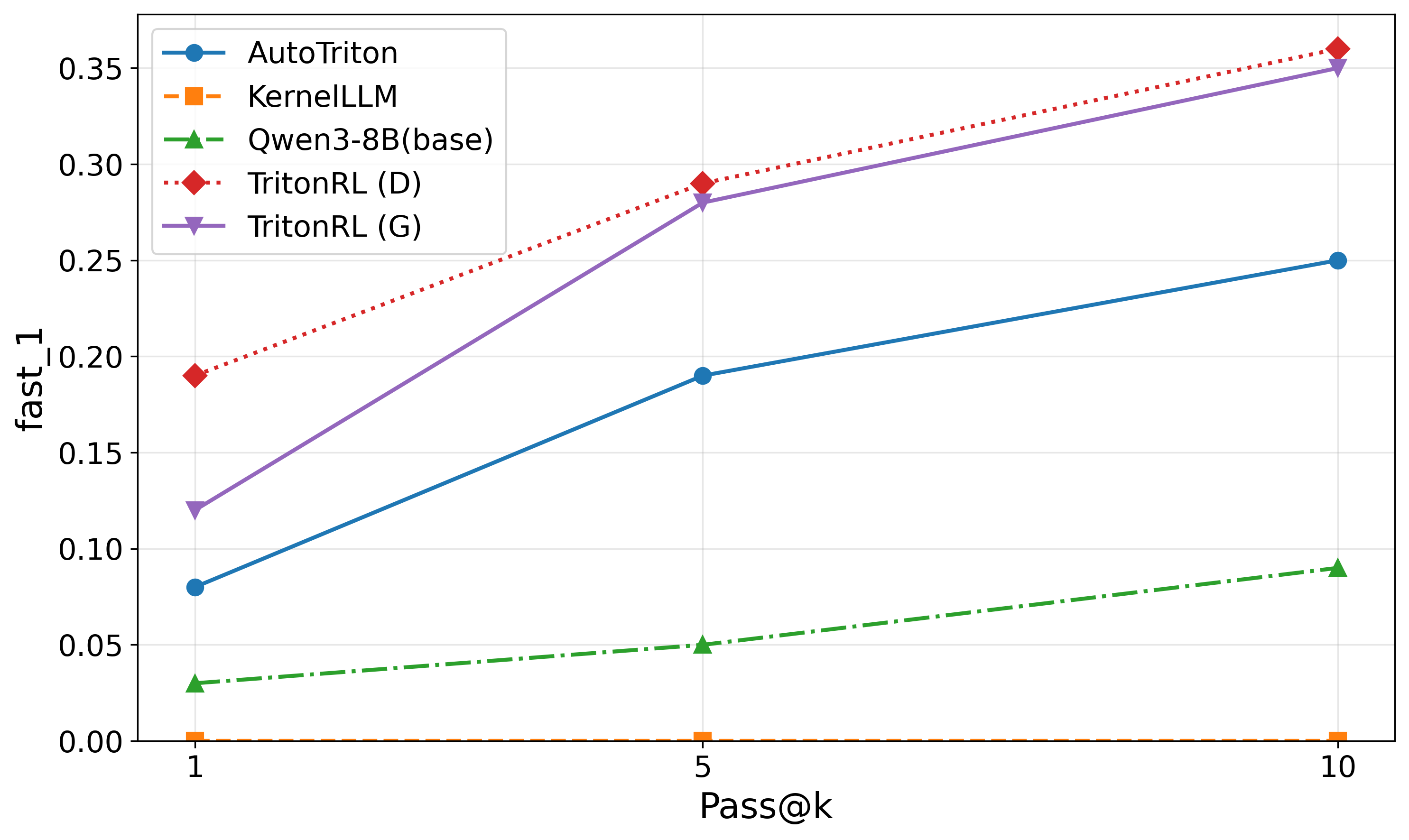}
  \caption{Pass@$k$ correctness and $\text{fast}_1$ for $k=1,5,10$ on KernelBench Level 1 tasks. 
The figures show how performance (correctness on the left, $\text{fast}_1$ on the right) of \ours and baseline models scales as the number of attempts increases.
  }
  \label{fig:level1_passk}
\end{figure}

\begin{figure}[h]
  \centering
  \includegraphics[width=0.47\linewidth]{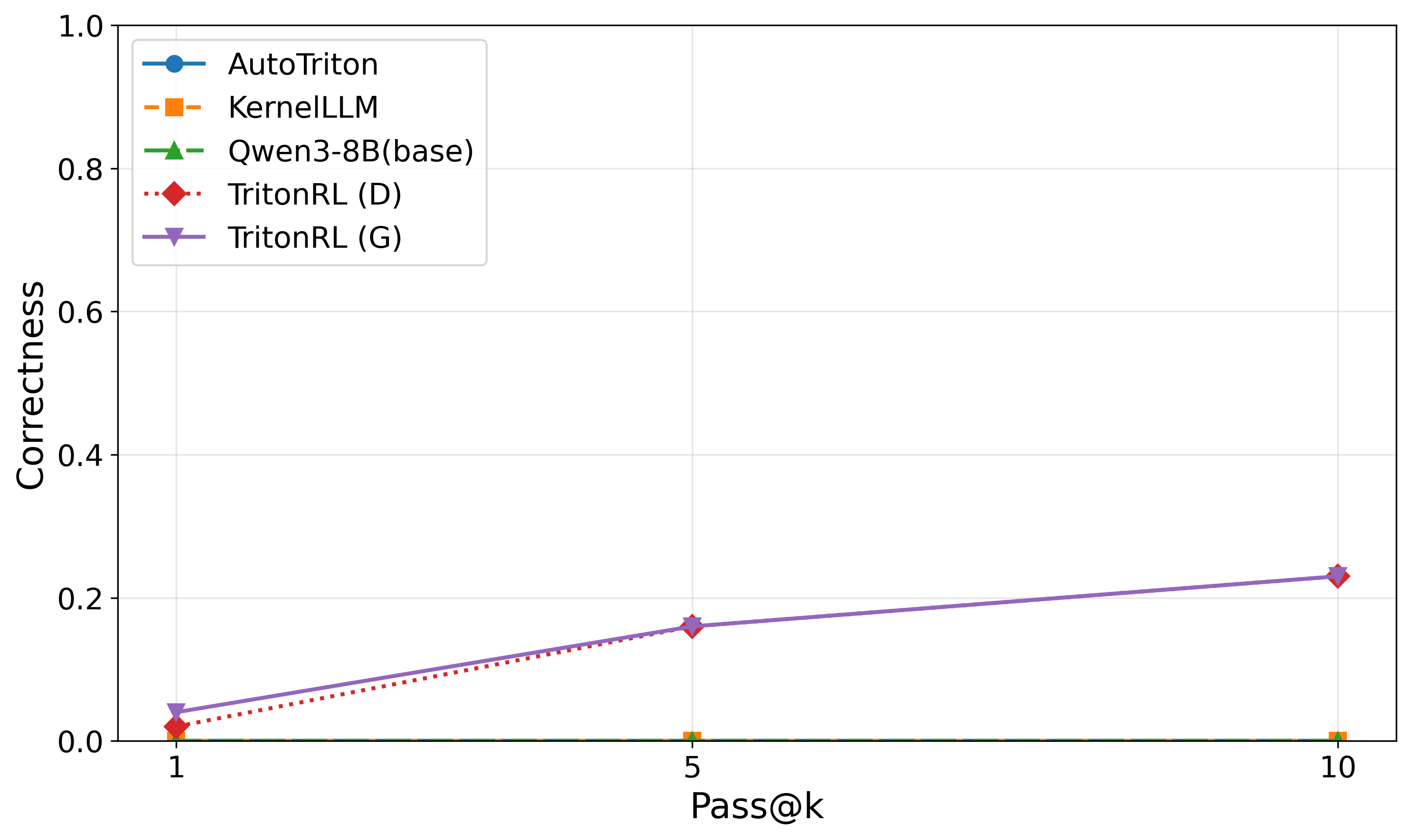}
  \hspace{0.1cm}
  \includegraphics[width=0.47\linewidth]{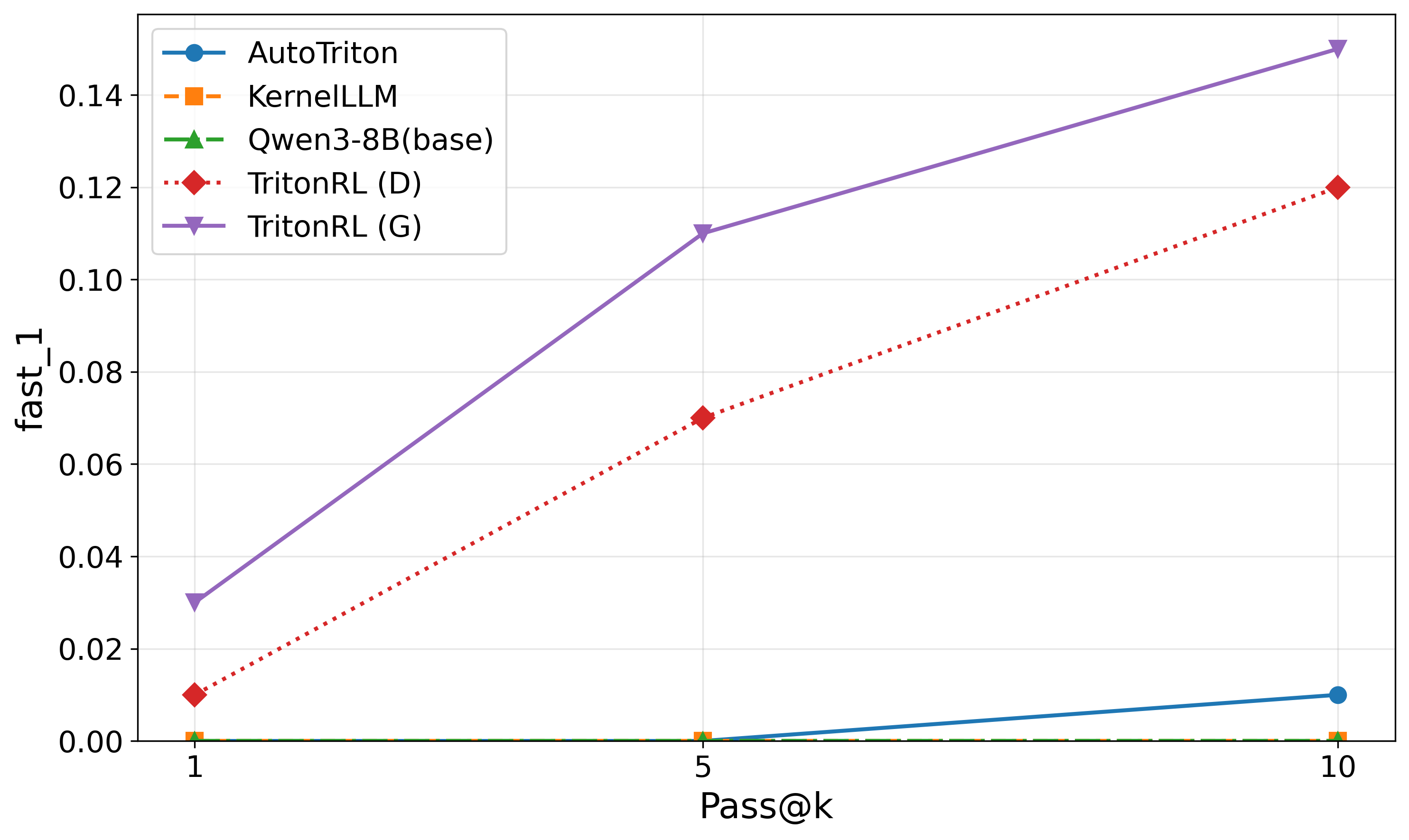}
  \caption{Pass@$k$ correctness and $\text{fast}_1$ for $k=1,5,10$ on KernelBench Level 2 tasks. 
The figures show how performance (correctness on the left, $\text{fast}_1$ on the right) of \ours and baseline models scales as the number of attempts increases.
  }
  \label{fig:level2_passk}
\end{figure}

We evaluated inference scaling by varying the number of sampled attempts ($k=1, 5, 10$), as illustrated in Figure~\ref{fig:level1_passk} and Figure~\ref{fig:level2_passk}. The correctness and runtime efficiency of \ours increases with more samples, whereas KernelLLM and Qwen3-8B show limited improvement, suggesting that \ours generates a more diverse set of codes and benefits from additional sampling during inference. %

We also report pass@1 and pass@5 results for KernelBench Level 1 and Level 2 tasks in Table~\ref{tab:kernelbench_passk_level1} and Table~\ref{tab:kernelbench_passk_level2}. These results further demonstrate the strong performance of \ours in generating valid, correct, and efficient Triton code across various pass@$k$ metrics.

\begin{table*}[ht!] \small
  \caption{Pass@k performance comparison for $k=1, 5, 10$ on KernelBench Level 1 tasks.
  }
  \label{tab:kernelbench_passk_level1}
  \centering
  \resizebox{0.99\textwidth}{!}{
    \begin{tabular}{lccccccccccc}
      \toprule
      \multirow{2}{*}{\textbf{Model}} 
      & \multicolumn{3}{c}{\textbf{\textsc{pass@1}}} & \multicolumn{3}{c}{\textbf{\textsc{pass@5}}}  & \multicolumn{3}{c}{\textbf{\textsc{pass@10}}} \\
      \cmidrule(lr){2-4}
      \cmidrule(lr){5-7}
      \cmidrule(lr){8-10}
      & \textbf{valid} & \textbf{compiled}~/~\textbf{correct} & $\textbf{fast}_\textbf{1}$~/~$\textbf{fast}_\textbf{2}$ 
      & \textbf{valid} & \textbf{compiled}~/~\textbf{correct} & $\textbf{fast}_\textbf{1}$~/~$\textbf{fast}_\textbf{2}$ 
      & \textbf{valid} & \textbf{compiled}~/~\textbf{correct} & $\textbf{fast}_\textbf{1}$~/~$\textbf{fast}_\textbf{2}$ 
      \\         
      \midrule
        Qwen3 (8B) & $70.0$ & $65.0$~/~$10.0$ & $3.0$~/~$2.0$ & $96.0$ & $96.0$~/~$17.0$ & $5.0$~/~$4.0$ & $99.0$  & $99.0$~/~$23.0$ & $9.0$~/~$5.0$ \\
        Qwen3 (14B) & $80.0$ & $73.0$~/~$17.0$ & $4.0$~/~$2.0$ & $99.0$ & $96.0$~/~$23.0$ & $7.0$~/~$6.0$ & $100.0$ &$99.0$~/~$35.0$ & $9.0$~/~$7.0$  \\
        Qwen3 (32B) & $81.0$ & $73.0$~/~$27.0$ & $7.0$~/~$5.0$ & $99.0$ & $96.0$~/~$53.0$ & $32.0$~/~$11.0$ & $100.0$ & $100.0$~/~$62.0$  &$39.0$~/~$16.0$    \\
         \midrule
      KernelLLM &  $32.0$ & $29.0$~/~$14.0$ & $0.0$~/~$0.0$ & $41.0$ & $39.0$~/~$19.0$ & $0.0$~/~$0.0$ & $42.0$  & $40.0$~/~$20.0$ & $0.0$~/~$0.0$ \\
      AutoTriton & $65.0$ & $57.0$~/~$35.0$ & $8.0$~/~$4.0$ & $86.0$ & $80.0$~/~$54.0$ & $19.0$~/~$7.0$ & $97.0$  & $92.0$~/~$57.0$ & $25.0$~/~$10.0$  \\
      \ours (D)  & $88.0$  & $84.0$~/~$44.0$ & $19.0$~/~$8.0$ & $100.0$ & $100.0$~/~$70.0$ & $29.0$~/~$12.0$ & $100.0$    & $100.0$~/~$78.0$     &  $36.0$~/~$13.0$   \\
      \ours (D) + IA & $88.0$  & $83.0$~/~$37.0$ & $12.0$~/~$9.0$ & $100.0$ & $99.0$~/~$69.0$ & $28.0$~/~$14.0$ & $100.0$ & $99.0$~/~$78.0$ & $34.0$~/~$15.0$\\ 
      \ \ \ SFT w. DeepSeek-R1 (w/o RL) & $75.0$  & $70.0$~/$31.0$ & $13.0$~/~$7.0$ & $99.0$ & $98.0$~/~$49.0$ & $25.0$~/$10.0$ &  $100.0$ & $100.0$~/~$54.0$   & $28.0$~/~$12.0$  \\ 
      \ours (G)   &  $69.0$ & $64.0$~/~$25.0$ & $12.0$~/~$9.0$ & $100.0$ & $98.0$~/~$59.0$ & $28.0$~/~$15.0$ & $100.0$ & $100.0$~/~$76.0$ & $35.0$~/$17.0$ \\
      \ours (G) + IA  & $72.0$  & $72.0$~/~$29.0$ & $13.0$~/~$10.0$ & $100.0$ & $100.0$~/~$77.0$ & $36.0$~/~$22.0$ & $100.0$    & $100.0$~/~$88.0$     &  $41.0$~/~$22.0$   \\
      \ \ \ SFT w. GPT-OSS 120B (w/o RL) & $43.0$ & $43.0$~/~$15.0$ & $6.0$~/~$5.0$ & $94.0$ & $92.0$~/~$48.0$ & $19.0$~/~$15.0$ & $100.0$ & $100.0$~/~$64.0$ & $26.0$~/~$18.0$ \\ 
      \midrule \midrule
        Claude-3.7 & $79.0$ & $74.0$~/~$31.0$ & $10.0$~/~$5.0$ & $100.0$ & $100.0$~/~$51.0$ & $21.0$~/~$11.0$ & $100.0$ & $100.0$~/~$63.0$ & $29.0$~/~$16.0$  \\
        DeepSeek-R1 ($685$B) &  $83.0$ & $81.0$~/~$32.0$ & $13.0$~/~$5.0$ & $99.0$ & $98.0$~/~$56.0$ & $25.0$~/~$12.0$ & $100.0$ & $100.0$~/~$69.0$ & $28.0$~/~$13.0$ \\
      GPT-oss ($120$B) & $79.0$ & $78.0$~/~$46.0$ & $18.0$~/~$10.0$ & $100.0$ & $100.0$~/~$69.0$ & $30.0$~/~$18.0$ & $100.0$ & $100.0$~/~$81.0$ & $33.0$~/~$22.0$ \\
        \bottomrule
  \end{tabular}
  }
\end{table*}

\begin{table*}[ht!] \small
  \caption{Pass@k performance comparison for $k=1, 5, 10$ on KernelBench Level 2 tasks.
  }
  \label{tab:kernelbench_passk_level2}  
  \centering
  \resizebox{0.99\textwidth}{!}{
    \begin{tabular}{lccccccccccc}
      \toprule
      \multirow{2}{*}{\textbf{Model}} 
      & \multicolumn{3}{c}{\textbf{\textsc{pass@1}}} & \multicolumn{3}{c}{\textbf{\textsc{pass@5}}}  & \multicolumn{3}{c}{\textbf{\textsc{pass@10}}} \\
      \cmidrule(lr){2-4}
      \cmidrule(lr){5-7}
      \cmidrule(lr){8-10}
      & \textbf{valid} & \textbf{compiled}~/~\textbf{correct} & $\textbf{fast}_\textbf{1}$~/~$\textbf{fast}_\textbf{2}$ 
      & \textbf{valid} & \textbf{compiled}~/~\textbf{correct} & $\textbf{fast}_\textbf{1}$~/~$\textbf{fast}_\textbf{2}$ 
      & \textbf{valid} & \textbf{compiled}~/~\textbf{correct} & $\textbf{fast}_\textbf{1}$~/~$\textbf{fast}_\textbf{2}$ 
      \\         
      \midrule
        Qwen3 (8B) & $7.0$ & $7.0$~/~$0.0$ & $0.0$~/~$0.0$ & $29.0$ & $29.0$~/~$0.0$ & $0.0$~/~$0.0$ & $38.0$ & $37.0$~/~$1.0$ & $0.0$~/~$0.0$ \\
        Qwen3 (14B) & $5.0$ & $5.0$~/~$0.0$ & $0.0$~/~$0.0$ & $26.0$ & $24.0$~/~$0.0$ & $0.0$~/~$0.0$ & $40.0$ & $38.0$~/~$1.0$ & $0.0$~/~$0.0$ \\
        Qwen3 (32B) & $10.0$ & $10.0$~/~$1.0$ & $1.0$~/~$1.0$ & $28.0$ & $26.0$~/~$2.0$ & $2.0$~/~$1.0$ & $36.0$ & $35.0$~/~$2.0$ & $2.0$~/~$1.0$ \\ \midrule
        KernelLLM & $0.0$ & $0.0$~/~$0.0$ & $0.0$~/~$0.0$ & $1.0$ & $0.0$~/~$0.0$ & $0.0$~/~$0.0$  & $3.0$ & $0.0$~/~$0.0$ & $0.0$~/~$0.0$  \\
        AutoTriton & $1.0$ & $1.0$~/~$0.0$ & $0.0$~/~$0.0$ & $5.0$ & $5.0$~/~$0.0$ & $0.0$~/~$0.0$  & $11.0$ & $10.0$~/~$1.0$ & $0.0$~/~$0.0$\\
        \ours (D)  &  $28.0$ & $27.0$~/~$2.0$ & $1.0$~/~$0.0$ & $55.0$ & $55.0$~/~$16.0$ & $7.0$~/~$1.0$  & $63.0$ & $63.0$~/~$23.0$ & $12.0$~/~$3.0$  \\
        \ours (D) + IA & $24.0$ & $24.0$~/~$3.0$ & $3.0$~/~$2.0$    & $42.0$ & $41.0$~/~$14.0$ & $10.0$~/~$5.0$   & $52.0$ & $51.0$~/~$24.0$ & $16.0$~/~$7.0$  \\
        \ \ \ SFT w. DeepSeek-R1 (w/o RL) & $22.0$  & $19.0$~/~$4.0$ & $4.0$~/~$0.0$ & $50.0$  & $45.0$~/~$10.0$ & $9.0$~/~$2.0$ & $55.0$  & $51.0$~/~$13.0$ & $10.0$~/~$3.0$ & \\ 
        \ours (G)   &  $49.0$ & $47.0$~/~$4.0$ & $3.0$~/~$1.0$ & $88.0$ & $87.0$~/~$16.0$ & $11.0$~/~$2.0$  & $91.0$ & $91.0$~/~$23.0$ & $15.0$~/~$7.0$ \\
        \ours (G) + IA   &  $44.0$ & $44.0$~/~$6.0$ & $5.0$~/~$2.0$ & $83.0$ & $82.0$~/~$18.0$ & $14.0$~/~$6.0$  & $93.0$ & $93.0$~/~$28.0$ & $19.0$~/~$10.0$  \\
        \ \ \ SFT w. GPT-OSS 120B (w/o RL)  & $41.0$  & $40.0$~/~$4.0$ & $4.0$~/~$0.0$  & $80.0$  & $79.0$~/~$10.0$ & $7.0$~/~$1.0$ & $88.0$  & $88.0$~/~$13.0$ & $9.0$~/~$3.0$  \\         
        \midrule \midrule
        Claude-3.7  & $16.0$ & $13.0$~/~$2.0$ & $0.0$~/~$0.0$ & $32.0$ & $31.0$~/~$12.0$ & $3.0$~/~$0.0$ &  $34.0$  & $34.0$~/~$14.0$ & $4.0$~/~$1.0$ \\
        DeepSeek-R1& $7.0$ & $5.0$~/~$1.0$ & $0.0$~/~$0.0$ & $24.0$ & $23.0$~/~$7.0$ & $4.0$~/~$3.0$ &  $31.0$ &  $31.0$~/~$14.0$  & $9.0$~/~$4.0$  \\
        GPT-oss ($120$B) & $9.0$ & $9.0$~/~$2.0$ & $1.0$~/~$1.0$ & $30.0$ & $29.0$~/~$9.0$ & $6.0$~/~$2.0$ & $39.0$ & $39.0$~/~$15.0$ & $12.0$~/~$4.0$  \\
        \bottomrule
  \end{tabular}
  }
\end{table*}

\subsection{Robustness gains from diverse input augmentation (IA)}
\label{sec:input_augmentation_ablation}
To demonstrate the effectiveness of our input augmentation (IA) strategy in enhancing model robustness to diverse input shapes, we conducted additional experiments on input-augmented KernelBench Level 1 and Level 2 tasks, where five distinct input shapes were used for each of the 100 tasks per level (500 total evaluations per level). The results, presented in Table~\ref{tab:kernelbench_diverseinput}, indicate IA improves correctness and speedup for small $k$ (pass@1 and pass@5), showing its role in enhancing robustness to input variations on initial attempts. As $k$ increases to 10, the performance gap narrows or even reverses, suggesting that with sufficient sampling models can produce high-quality outputs regardless of exposure to diverse input shapes during training. Nonetheless, IA remains beneficial for robustness, particularly in low-sample scenarios.

\begin{table*}[ht!] \small
  \caption{Pass@k results (k = 1, 5, 10) on input-augmented KernelBench Level 1 and 2. Each of 100 tasks for each level was evaluated with five distinct input shapes (500 total for each level).}
  \label{tab:kernelbench_diverseinput}  
  \centering
    \resizebox{0.99\textwidth}{!}{
      \begin{tabular}{lccccccccccc}
        \toprule
        \multicolumn{10}{c}{\textbf{Level 1}} \\
        \cmidrule(lr){2-10}
        \multirow{2}{*}{\textbf{Model}} 
        & \multicolumn{3}{c}{\textbf{\textsc{pass@1}}} & \multicolumn{3}{c}{\textbf{\textsc{pass@5}}}  & \multicolumn{3}{c}{\textbf{\textsc{pass@10}}} \\
        \cmidrule(lr){2-4}
        \cmidrule(lr){5-7}
        \cmidrule(lr){8-10}
      & \textbf{valid} & \textbf{compiled}~/~\textbf{correct} & $\textbf{fast}_\textbf{1}$~/~$\textbf{fast}_\textbf{2}$ 
      & \textbf{valid} & \textbf{compiled}~/~\textbf{correct} & $\textbf{fast}_\textbf{1}$~/~$\textbf{fast}_\textbf{2}$ 
      & \textbf{valid} & \textbf{compiled}~/~\textbf{correct} & $\textbf{fast}_\textbf{1}$~/~$\textbf{fast}_\textbf{2}$ 
      \\         
      \midrule
        \ours (D)  & $88.0$ & $85.0$~/~$37.0$ & $12.0$~/~$6.0$     & $100.0$ & $100.0$~/$64.0$ & $27.0$~/~$10.0$  & $100.0$ & $100.0$~/~$75.0$ & $33.0$~/~$12.0$ \\
        \ours (D) + IA & $85.0$ & $81.0$~/~$39.0$ & $14.0$~/~$7.0$       & $100.0$ & $99.0$~/~$65.0$ & $27.0$~/~$11.0$ & $100.0$ & $99.0$~/~$72.0$ & $32.0$~/~$13.0$ \\
        \ours (G)   & $64.0$ & $62.0$~/~$30.0$ & $12.0$~/~$7.0$ & $99.0$ & $99.0$~/~$67.0$ & $26.0$~/~$17.0$ & $100.0$ & $100.0$~/~$80.0$ & $31.0$~/~$19.0$ \\
        \ours (G) + IA   & $77.0$ & $77.0$~/~$38.0$ & $14.0$~/~$8.0$    & $99.0$ & $99.0$~/~$73.0$ & $29.0$~/~$18.0$ & $100.0$ & $100.0$~/~$83.0$ & $36.0$~/~$20.0$\\
        \midrule
        \midrule
        \multicolumn{10}{c}{\textbf{Level 2}} \\
        \cmidrule(lr){2-10}
        \multirow{2}{*}{\textbf{Model}} 
        & \multicolumn{3}{c}{\textbf{\textsc{pass@1}}} & \multicolumn{3}{c}{\textbf{\textsc{pass@5}}}  & \multicolumn{3}{c}{\textbf{\textsc{pass@10}}} \\
        \cmidrule(lr){2-4}
        \cmidrule(lr){5-7}
        \cmidrule(lr){8-10}
      & \textbf{valid} & \textbf{compiled}~/~\textbf{correct} & $\textbf{fast}_\textbf{1}$~/~$\textbf{fast}_\textbf{2}$ 
      & \textbf{valid} & \textbf{compiled}~/~\textbf{correct} & $\textbf{fast}_\textbf{1}$~/~$\textbf{fast}_\textbf{2}$ 
      & \textbf{valid} & \textbf{compiled}~/~\textbf{correct} & $\textbf{fast}_\textbf{1}$~/~$\textbf{fast}_\textbf{2}$ 
      \\         
      \midrule
        \ours (D)  & $31.0$ & $29.0$~/~$5.0$ & $2.0$~/~$1.0$         & $56.0$ & $54.0$~/~$18.0$ & $9.0$~/~$3.0$     & $66.0$ & $65.0$~/~$27.0$ & $15.0$~/~$5.0$ \\
        \ours (D) + IA & $25.0$ & $24.0$~/~$5.0$ & $3.0$~/~$1.0$         & $43.0$ & $42.0$~/~$13.0$ & $8.0$~/~$3.0$     & $52.0$ & $51.0$~/~$20.0$ & $13.0$~/~$4.0$\\
        \ours (G)   & $44.0$ & $42.0$~/~$5.0$ & $3.0$~/~$1.0$     & $85.0$ & $84.0$~/~$15.0$ & $11.0$~/~$5.0$     & $94.0$ & $93.0$~/~$23.0$ & $14.0$~/~$6.0$    \\
        \ours (G) + IA   & $56.0$ & $54.0$~/~$8.0$ & $6.0$~/~$3.0$       & $88.0$ & $86.0$~/~$22.0$ & $14.0$~/~$7.0$       & $94.0$ & $93.0$~/~$29.0$ & $19.0$~/~$10.0$ \\
        \bottomrule
      \end{tabular}
  }
\end{table*}

\subsection{Ablation on base model choices}
\label{sec:base_model_ablation}
In this section, we investigate the impact of different base models for fine-tuning with our RL framework. We compare two base models: Qwen3-8B, which is our main base model used throughout this paper, and Seed-Coder-8B-Reasoning~\citep{zhang2025seed}, which is the base model of AutoTriton \citep{li2025autotriton}. As shown in Table~\ref{tab:basemodel_comparison}, \ours fine-tuned from Seed-Coder-8B-Reasoning outperforms AutoTriton, which is also fine-tuned from the same base model, across all metrics on KernelBench Level 1 and Level 2 tasks. This demonstrates the effectiveness of our RL fine-tuning approach regardless of the base model. Furthermore, when comparing \ours fine-tuned from Qwen3-8B and from Seed-Coder-8B-Reasoning, we observe that the former achieves significantly higher correctness and speedup on both Level 1 and Level 2 tasks, indicating that Qwen3-8B is a more effective foundation for our RL fine-tuning.

\begin{table*}[ht!] \small
    \caption{Comparison of \ours fine-tuned from different base models (Qwen3-8B and Seed-Coder-8B-Reasoning) on KernelBench Level 1 and Level 2. 
      All metrics are reported as pass@10 (\%). \ours (G) indicates our model initialized from the GPT-OSS 120B SFT checkpoint; + IA denotes input augmentation. 
      "Qwen3-8B + \ours (G) + IA" and "Seed-Coder-8B-Reasoning + \ours (G) + IA" refer to models fine-tuned with our RL framework from the respective base models under identical settings. 
      AutoTriton uses Seed-Coder-8B-Reasoning as its base model and can be directly compared to "Seed-Coder-8B-Reasoning + \ours (G) + IA".
    }
    \label{tab:basemodel_comparison}
  \centering
    \resizebox{0.98\textwidth}{!}{
        \begin{tabular}{lccccccc}
        \toprule
        \multirow{2}{*}{\textbf{Model}} 
        & \multirow{2}{*}{\textbf{\#Params}} 
        & \multicolumn{3}{c}{\textbf{\textsc{Level1}}} 
        & \multicolumn{3}{c}{\textbf{\textsc{Level2}}} \\
        \cmidrule(lr){3-5}
        \cmidrule(lr){6-8}
        & & \textbf{valid} & \textbf{compiled}~/~\textbf{correct} 
        & $\textbf{fast}_1$~/~$\textbf{fast}_2$ & \textbf{valid} 
        & \textbf{compiled}~/~\textbf{correct} & $\textbf{fast}_1$~/~$\textbf{fast}_2$ 
        \\         
        \midrule
        Seed-Coder-8B-Reasoning (base)  & $8$B  &  $91.0$ & $82.0$~/~$19.0$ & $7.0$~/~$6.0$  & $49.0$ & $12.0$~/~$1.0$  & $1.0$~/~$0.0$ \\
        AutoTriton & $8$B &$97.0$  & $92.0$~/~$57.0$ & $25.0$~/~$10.0$ & $11.0$ & $10.0$~/~$1.0$ & $0.0$~/~$0.0$   \\
        Seed-Coder-8B-Reasoning + \ours (G) + IA  & $8$B  &  $\textbf{99.0}$ & $\textbf{99.0}$~/~$\textbf{61.0}$ & $\textbf{27.0}$~/~$\textbf{17.0}$  &  $\textbf{86.0}$ & $\textbf{85.0}$~/~$\textbf{17.0}$ & $\textbf{11.0}$~/~$\textbf{5.0}$   \\
        \midrule \midrule
        Qwen3-8B (base)  &   $8$B    & $99.0$ & $99.0$~/~$23.0$ & $9.0$~/~$5.0$ & $38.0$ & $37.0$~/~$0.0$ & $0.0$~/~$0.0$ \\
        Qwen3-8B + \ours (G) + IA  & $8$B  &   $100.0$    & $100.0$~/~$88.0$     &  $41.0$~/~$22.0$ &  $93.0$ & $93.0$~/~$28.0$   &  $19.0$~/~$10.0$ \\
        \bottomrule
    \end{tabular}
    }
    \vspace{-3mm}
\end{table*}

\subsection{Comparison of token-class reward assignments}
\label{sec:reward_assignment_ablation}

In GRPO, the advantage for the "code" component is computed relative to other codes sampled in the same group, but those codes may be paired with different prior reasoning traces (plans). 
Penalizing a correct implementation solely because it was conditioned on a weak plan is undesirable: it risks discouraging valid Triton implementations and degrading the base model’s already limited Triton skills. Thus, we assign correctness-based rewards for code tokens to avoid such unwarranted penalties and to preserve model's capability to implement correct Triton code.

We compare our choice of reward assignment (defined in \eqref{eq:reward}),
\begin{align}
r_{i,j}^{\text{\tiny plan}} = R^{\text{speedup}}(g_i, o_{i,j}), \quad
r_{i,j}^{\text{\tiny code}} = R^{\text{correct}}(g_i, o_{i,j}),
\end{align}
against models trained using the intuitive speedup rewards for both plan and code tokens, i.e.,
\begin{align}\label{eq:reward-speedup-speedup}
r_{i,j}^{\text{\tiny plan}} = R^{\text{speedup}}(g_i, o_{i,j}), \quad
r_{i,j}^{\text{\tiny code}} = R^{\text{speedup}}(g_i, o_{i,j}),
\end{align}
for the same $\alpha=0.1,$ starting from the same SFT base model distilled from DeepSeek-R1. We denote our default reward assignment (\eqref{eq:reward}) as speedup-correct and the latter (\eqref{eq:reward-speedup-speedup}) as speedup-speedup assignment in the Table~\ref{tab:reward_assignment_ablation}.

As shown in Table~\ref{tab:reward_assignment_ablation}, giving correctness feedback to code tokens yields better performance in both speedup and correctness than giving speedup-based feedback to all tokens. As explained above, correctness rewards for code tokens prevent accurate implementations conditioned on suboptimal plans from being unfairly penalized, ultimately helping the model learn to generate kernels that are more efficient.

\begin{table*}[ht!] \small
\caption{ Comparison of token-class reward assignments (speedup-correct (ours) vs. speedup-speedup) on KernelBench Level 1 tasks. All metrics are reported as pass@10 (\%).}
    \vspace{2mm}
  \label{tab:reward_assignment_ablation}
  \centering
  \resizebox{0.55\textwidth}{!}{
    \begin{tabular}{lcccc}
    \toprule
    \multirow{1}{*}{\textbf{reward assignment type}} 
    & \textbf{valid} & \textbf{compiled}~/~\textbf{correct} & $\textbf{fast}_\textbf{1}$~/~$\textbf{fast}_\textbf{2}$ \\
    \midrule
    speedup-correct (ours) & $100.0$    & $100.0$~/~$78.0$     &  $36.0$~/~$13.0$   \\
    speedup-speedup & $100.0$    & $100.0$~/~$78.0$     &  $33.0$~/~$13.0$   \\
    \bottomrule
  \end{tabular}
  }
\end{table*}

\section{Data curation and examples}
\subsection{Data Mixing Subset Creation}
\label{sec:appendix:data-mixing-subset}
We labeled difficulty level of 11k PyTorch reference codes in KernelBook based on the complexity of kernel implementation using Qwen3-235B-Instruct \citep{qwen3tech}.
For each given PyTorch reference code, we prompt Qwen3-235B-Instruct (temperature=0.7, top\_p=0.8) to label the difficulty level of replacing the PyTorch reference with Triton code as follows:
\begin{tcolorbox}[colback=gray!10, colframe=gray!80, title=Instruction (input) example]

\texttt{``` <PyTorch reference code> ```}

    Assign a kernel implementation complexity level (1, 2, or 3) of the provided reference PyTorch architecture according to the criteria below:

    • Level 1: Single primitive operation. This level includes the foundational building blocks
    of AI (e.g. convolutions, matrix-vector and matrix-matrix multiplications, losses, activations, and layer
    normalizations).
    Since PyTorch makes calls to several well-optimized and often closed-source kernels under-the-hood, it
    can be challenging for LMs to outperform the baseline for these primitive operations. However, if an LM
    succeeds, the open-source kernels could be an impactful alternative to the closed-source (e.g., CuBLAS [27])
    kernels.

    • Level 2: Operator sequences. This level includes AI workloads containing multiple
    primitive operations, which can be fused into a single kernel for improved performance (e.g., a combination
    of a convolution, ReLU, and bias).
    Since compiler-based tools such as the PyTorch compiler are effective at fusion, it can be challenging for
    LMs to outperform them. However, LMs may propose more complex algorithms compared to compiler
    rules.

    • Level 3: This level includes architectures that power popular AI
    models, such as AlexNet and MiniGPT, collected from popular PyTorch repositories on GitHub.
\end{tcolorbox}

\subsection{Example of invalid Triton codes in KernelBook}
\label{sec:invalid_kernelbook_example}

The following is an example of invalid Triton code from KernelBook. The intended objective is to implement a convolution layer with bias addition using Triton. However, the generated code fails to correctly realize the convolution operation and bias addition within the Triton kernel, resulting in incorrect functionality.
Specifically, this is not a valid Triton implementation of conv2d because the actual convolution is performed outside the Triton kernel using PyTorch modules. The Triton kernel itself only adds a bias term to the convolution output by loading and adding per-channel bias values; it does not perform any of the core convolution computations, such as sliding window access, dot products, or handling stride, padding, or dilation. Thus, while the Triton code is syntactically valid and functionally correct for bias addition, it does not implement conv2d and cannot be considered a genuine Triton-based convolution.

\begin{lstlisting}[
  language=Python,
  basicstyle=\ttfamily\footnotesize,
  frame=single,
  breaklines=true,
  columns=fullflexible
]
import torch
import torch.nn as nn
import triton
import triton.language as tl

from torch._inductor.select_algorithm import extern_kernels
from torch._inductor.runtime.triton_heuristics import grid
from torch._C import _cuda_getCurrentRawStream as get_raw_stream
from torch._C._dynamo.guards import assert_size_stride


@triton.jit
def triton_poi_fused_convolution_0(in_out_ptr0, in_ptr0, xnumel, XBLOCK: tl.constexpr):
    xoffset = tl.program_id(0) * XBLOCK
    xindex = xoffset + tl.arange(0, XBLOCK)[:]
    
    tl.full([XBLOCK], True, tl.int1)

    x3 = xindex
    x1 = xindex // 4096 %

    tmp0 = tl.load(in_out_ptr0 + x3, None)
    tmp1 = tl.load(in_ptr0 + x1, None, eviction_policy='evict_last')

    tmp2 = tmp0 + tmp1

    tl.store(in_out_ptr0 + x3, tmp2, None)


def call(args):
    primals_1, primals_2, primals_3 = args
    args.clear()

    assert_size_stride(primals_1, (2, 3, 1, 1), (3, 1, 1, 1))
    assert_size_stride(primals_2, (2,), (1,))
    assert_size_stride(primals_3, (4, 3, 64, 64), (12288, 4096, 64, 1))

    with torch.cuda._DeviceGuard(0):
        torch.cuda.set_device(0)

        buf0 = extern_kernels.convolution(
            primals_3,
            primals_1,
            stride=(1, 1),
            padding=(0, 0),
            dilation=(1, 1),
            transposed=False,
            output_padding=(0, 0),
            groups=1,
            bias=None
        )

        assert_size_stride(buf0, (4, 2, 64, 64), (8192, 4096, 64, 1))

        buf1 = buf0
        del buf0

        get_raw_stream(0)

        triton_poi_fused_convolution_0[grid(32768)](
            buf1, primals_2, 32768,
            XBLOCK=128,
            num_warps=4,
            num_stages=1
        )

        del primals_2

    return buf1, primals_1, primals_3


class ModelNew(nn.Module):
    def __init__(self, num_classes=2, in_channels=3, is_deconv=False, is_batchnorm=False, *args, **kwargs):
        super(ModelNew, self).__init__()
        self.is_deconv = is_deconv
        self.in_channels = in_channels
        self.is_batchnorm = is_batchnorm

        self.final = nn.Conv2d(self.in_channels, num_classes, kernel_size=1)

    def forward(self, input_0):
        primals_1 = self.final.weight
        primals_2 = self.final.bias
        primals_3 = input_0

        output = call([primals_1, primals_2, primals_3])
        return output[0]
\end{lstlisting}

\subsection{Example of invalid Triton codes for Level 2 tasks}
\label{sec:invalid_l2_example}

The following is an example of invalid Triton code generated during RL training of \ours. This level 2 task aims to implement a two-branch linear transformation with message passing, structured like a Graph Neural Network (GNN) layer. However, while the code implement linear transformations using a custom Triton kernel, it fails to fully realize the intended GNN functionality with only Triton kernels by using torch.bmm. 

\begin{lstlisting}[
  language=Python,
  basicstyle=\ttfamily\footnotesize,
  frame=single,
  breaklines=true,
  columns=fullflexible
]
import torch
import torch.nn as nn
import triton
import triton.language as tl

@triton.jit
def _linear_triton_kernel(
    input_ptr, weight_ptr, bias_ptr, output_ptr,
    B, M, N, K,
    stride_input_b, stride_input_m, stride_input_k,
    stride_weight_n, stride_weight_k,
    stride_output_b, stride_output_m, stride_output_n,
    BLOCK_SIZE_M: tl.constexpr, BLOCK_SIZE_N: tl.constexpr, BLOCK_SIZE_K: tl.constexpr,
):
    pid_b = tl.program_id(0)
    pid_m = tl.program_id(1)
    pid_n = tl.program_id(2)

    pid_m = tl.multiple_of(pid_m * BLOCK_SIZE_M, BLOCK_SIZE_M)
    pid_n = tl.multiple_of(pid_n * BLOCK_SIZE_N, BLOCK_SIZE_N)

    input_ptr += pid_b * stride_input_b
    output_ptr += pid_b * stride_output_b

    offs_m = pid_m + tl.arange(0, BLOCK_SIZE_M)
    offs_n = pid_n + tl.arange(0, BLOCK_SIZE_N)
    offs_k = tl.arange(0, BLOCK_SIZE_K)

    input_ptrs = input_ptr + (offs_m[:, None] * stride_input_m + offs_k[None, :] * stride_input_k)
    weight_ptrs = weight_ptr + (offs_n[:, None] * stride_weight_n + offs_k[None, :] * stride_weight_k)

    acc = tl.zeros((BLOCK_SIZE_M, BLOCK_SIZE_N), dtype=tl.float32)

    for k in range(0, K, BLOCK_SIZE_K):
        mask_k = (offs_k[None, :] < K - k)
        mask_m = (offs_m[:, None] < M)
        mask_n = (offs_n[None, :] < N)

        a = tl.load(input_ptrs, mask=mask_k & mask_m, other=0.0)
        b = tl.load(weight_ptrs, mask=mask_k & mask_n, other=0.0)

        b = tl.trans(b)
        acc += tl.dot(a, b)

        input_ptrs += BLOCK_SIZE_K * stride_input_k
        weight_ptrs += BLOCK_SIZE_K * stride_weight_k

    bias_ptrs = bias_ptr + offs_n
    bias = tl.load(bias_ptrs, mask=offs_n < N, other=0.0)
    acc += bias[None, :]

    offs_out_m = pid_m + tl.arange(0, BLOCK_SIZE_M)
    offs_out_n = pid_n + tl.arange(0, BLOCK_SIZE_N)
    out_ptrs = output_ptr + (offs_out_m[:, None] * stride_output_m + offs_out_n[None, :] * stride_output_n)

    mask = (offs_out_m[:, None] < M) & (offs_out_n[None, :] < N)
    tl.store(out_ptrs, acc, mask=mask)


def linear_triton(input, weight, bias):
    B, M, K = input.shape
    N = weight.shape[0]
    output = torch.empty((B, M, N), device=input.device, dtype=input.dtype)

    grid = (B, triton.cdiv(M, 64), triton.cdiv(N, 64))

    _linear_triton_kernel[grid](
        input, weight, bias, output,
        B, M, N, K,
        input.stride(0), input.stride(1), input.stride(2),
        weight.stride(0), weight.stride(1),
        output.stride(0), output.stride(1), output.stride(2),
        BLOCK_SIZE_M=64, BLOCK_SIZE_N=64, BLOCK_SIZE_K=64
    )

    return output


class ModelNew(nn.Module):
    def __init__(self, state_dim, name='', out_state_dim=None):
        super().__init__()
        self.state_dim = state_dim
        self.out_state_dim = out_state_dim if out_state_dim is not None else state_dim
        self.fc1 = nn.Linear(in_features=self.state_dim, out_features=self.out_state_dim)
        self.fc2 = nn.Linear(in_features=self.state_dim, out_features=self.out_state_dim)
        self.name = name

    def forward(self, input, adj):
        state_in = linear_triton(input, self.fc1.weight, self.fc1.bias)
        temp = torch.bmm(adj, input)
        forward_input = linear_triton(temp, self.fc2.weight, self.fc2.bias)
        return state_in + forward_input
\end{lstlisting}

\subsection{Invalid Triton code examples}
\label{sec:invalid_code_example}
Here, we provide some examples of invalid Triton code generated by the baseline models.

\paragraph{Example 1: Triton syntax failure (\texttt{syntax} = 0) -- no Triton kernel.}
This is not a Triton kernel because it does not include any custom \texttt{@triton.jit} kernel.
\begin{lstlisting}[
  language=Python,
  basicstyle=\ttfamily\footnotesize,
  frame=single,
  breaklines=true,
  columns=fullflexible
]
import torch
import torch.nn as nn
import triton
import triton.language as tl
import torch.nn.functional as F

# To replace with Triton, a custom kernel would be needed (not provided here)
class ModelNew(nn.Module):
    def __init__(self) -> None:
        super().__init__()

    def forward(self, x):
        return F.avg_pool2d(x, kernel_size=3, stride=None, padding=0)

def get_inputs():
    a = torch.randn(1, 128).cuda()
    b = torch.randn(1, 128).cuda()
    return [a, b]

def get_init_inputs():
    return []
\end{lstlisting}

\paragraph{Example 2: Functional failure (\texttt{syntax} = 1, \texttt{func} = 0) -- use of high-level PyTorch module.}
This implementation contains a Triton kernel, but the main computation (convolution) is still handled by the PyTorch \texttt{nn.Conv3d} module. The Triton kernel is only used for a simple addition, rather than implementing the full intended operation.
\begin{lstlisting}[
  language=Python,
  basicstyle=\ttfamily\footnotesize,
  frame=single,
  breaklines=true,
  columns=fullflexible
]
import torch
import torch.nn as nn
import triton
import triton.language as tl


@triton.jit
def add_bias_kernel(
    out_ptr,
    bias_ptr,
    n_elements,
    out_C,
    out_D,
    out_W,
    out_H,
    BLOCK_SIZE: tl.constexpr,
):
    pid = tl.program_id(0)
    block_start = pid * BLOCK_SIZE
    offsets = block_start + tl.arange(0, BLOCK_SIZE)
    mask = offsets < n_elements

    out = tl.load(out_ptr + offsets, mask=mask, other=0.0)

    stride_channel = out_D * out_W * out_H
    channel_idx = (offsets // stride_channel) %
    bias = tl.load(bias_ptr + channel_idx, mask=mask, other=0.0)

    out = out + bias
    tl.store(out_ptr + offsets, out, mask=mask)


def triton_add_bias(out: torch.Tensor, bias: torch.Tensor):
    assert out.is_cuda and bias.is_cuda, "Tensors must be on CUDA."

    out = out.contiguous()
    bias = bias.contiguous()
    n_elements = out.numel()
    BLOCK_SIZE = 128

    grid = lambda meta: (
        (n_elements + meta["BLOCK_SIZE"] - 1) // meta["BLOCK_SIZE"],
    )

    batch_size, out_channels, D, W, H = out.shape

    add_bias_kernel[grid](
        out,
        bias,
        n_elements,
        out_channels,
        D,
        W,
        H,
        BLOCK_SIZE=BLOCK_SIZE
    )

    return out


class ModelNew(nn.Module):
    def __init__(
        self,
        in_channels: int,
        out_channels: int,
        kernel_size: int,
        stride: int = 1,
        padding: int = 0,
        dilation: int = 1,
        groups: int = 1,
        bias: bool = False
    ):
        super(ModelNew, self).__init__()
        self.conv3d = nn.Conv3d(
            in_channels,
            out_channels,
            (kernel_size, kernel_size, kernel_size),
            stride=stride,
            padding=padding,
            dilation=dilation,
            groups=groups,
            bias=bias
        )

    def forward(self, x: torch.Tensor) -> torch.Tensor:
        out = self.conv3d(x)
        if self.conv3d.bias is not None:
            out = triton_add_bias(out, self.conv3d.bias)
        return out
\end{lstlisting}

\paragraph{Example 3: Functional failure (\texttt{syntax} = 1, \texttt{func} = 0) -- hardcoded output and no meaningful computation.}
While the Triton kernel is syntactically correct, but it doesn't actually implement the intended operation (Group Normalization). The kernel doesn't compute mean or variance, which are essential for GroupNorm. To implement real GroupNorm, you'd need to compute per-group statistics and normalize accordingly. Also, it only loads the input tensor and writes it back unchanged.
\begin{lstlisting}[
  language=Python,
  basicstyle=\ttfamily\footnotesize,
  frame=single,
  breaklines=true,
  columns=fullflexible
]
import torch
import torch.nn as nn
import triton
import triton.language as tl


@triton.jit
def groupnorm_kernel(
    x_ptr,  # Pointer to x tensor
    y_ptr,  # Pointer to y tensor (not used here)
    out_ptr,  # Pointer to output tensor
    n_elements,  # Total number of elements
    BLOCK_SIZE: tl.constexpr,
):
    # Each program handles a contiguous block of data of size BLOCK_SIZE
    block_start = tl.program_id(0) * BLOCK_SIZE
    # Create a range of offsets [0..BLOCK_SIZE-1]
    offsets = block_start + tl.arange(0, BLOCK_SIZE)
    # Mask to ensure we don't go out of bounds
    mask = offsets < n_elements
    # Load input value
    x = tl.load(x_ptr + offsets, mask=mask, other=0.0)
    # Compute x squared
    x_squared = x * x
    # Store the result
    tl.store(out_ptr + offsets, x, mask=mask)


def triton_groupnorm(x: torch.Tensor, y: torch.Tensor):
    assert x.is_cuda and y.is_cuda, "Tensors must be on CUDA."
    x = x.contiguous()
    y = y.contiguous()

    # Prepare output tensor
    out = torch.empty_like(x)

    # Number of elements in the tensor
    n_elements = x.numel()
    BLOCK_SIZE = 128  # Tunable parameter for block size

    # Determine the number of blocks needed
    grid = lambda meta: ((n_elements + meta["BLOCK_SIZE"] - 1) // meta["BLOCK_SIZE"],)

    # Launch the Triton kernel
    groupnorm_kernel[grid](x, y, out, n_elements, BLOCK_SIZE=BLOCK_SIZE)
    return out


class ModelNew(nn.Module):
    def __init__(self, num_features: int, num_groups: int) -> None:
        super().__init__()
        self.num_features = num_features
        self.num_groups = num_groups

    def forward(self, x: torch.Tensor) -> torch.Tensor:
        # Use Triton kernel for elementwise operations
        x_triton = triton_groupnorm(x, x)
        # Manually compute mean and variance (as Triton kernel only handles x)
        # Actual GroupNorm logic would go here
        # For this example, we return the Triton processed tensor
        return x_triton
\end{lstlisting}

\subsection{SFT and RL Dataset construction with KernelBook}
\label{sec:kernelbook_example}
To synthesize SFT dataset, we extract 11,621 PyTorch reference codes from KernelBook, executable without errors, such as
\begin{lstlisting}[
  language=Python,
  basicstyle=\ttfamily\footnotesize,
  frame=single,
  breaklines=true,
  columns=fullflexible
]
import torch
import torch.nn as nn


class Model(nn.Module):

def __init__(self):
super(Model, self).__init__()

def forward(self, neighbor):
return torch.sum(neighbor, dim=1)


def get_inputs():
return [torch.rand([4, 4, 4, 4])]


def get_init_inputs():
return [[], {}]
\end{lstlisting}

For each given PyTorch reference code, we construct an instruction for a teacher model to generate CoTs and Triton kernels as:
\begin{tcolorbox}[colback=gray!10, colframe=gray!80, title=Instruction (input) example]
Your task is to write custom Triton kernels to replace as many PyTorch operators as possible in the given architecture, aiming for maximum speedup. You may implement multiple custom kernels, explore operator fusion (such as combining matmul and relu), or introduce algorithmic improvements (like online softmax). You are only limited by your imagination.

You are given the following architecture:

\texttt{``` <PyTorch reference code> ```}

You have to optimize the architecture named Model with custom Triton kernels.
Optimize the architecture named Model with custom Triton kernels! Name your optimized output architecture ModelNew. Output the new code in codeblocks. Please generate real code, NOT pseudocode, make sure the code compiles and is fully functional. Just output the new model code, no other text, and NO testing code! Before writing a code, reflect on your idea to make sure that the implementation is correct and optimal.
\end{tcolorbox}

Given the instruction for each PyTorch reference code, we collect (CoT, Triton kernel code) pairs from DeepSeek-R1 and construct outputs for SFT by concatenating the pairs as follows:
\begin{tcolorbox}[colback=gray!10, colframe=gray!80, title=Triton kernel with CoT (output) example]
  \texttt{
    <think> \\
    CoT \\
    </think>
    \\
    ``` \\
    <Triton kernel code> \\
    ```
  }
\end{tcolorbox}

For RL training, we use the same instruction input as a prompt for the same set of Pytorch reference codes in KernelBook, without the output synthesized by DeepSeek-R1 because RL training only requires reward feedback, which can be directly obtained from executing the generated Triton code.

\section{LLM judge for functional validity verification}
\label{sec:appendix:cheating_detector}
To further enhance the functional validity verifier beyond unit tests, we employ an LLM-based judge to identify potential cheating behaviors in the generated Triton code, which may pass unit tests and syntax checks but fail to implement the intended functionality. We utilize Qwen3-235B-Instruct \citep{qwen3tech} as the judge model, prompted to determine if the Triton code genuinely implements the same functionality without relying on high-level PyTorch constructs or hardcoded constants.
We design the following prompt that details the functional requirements of Triton kernels for the judge model:
\begin{tcolorbox}[colback=gray!10, colframe=gray!80, title=Judge Instruction for Functional Validity Verification]
Detect whether the following Triton kernel implementation of an operation is cheating or not. A kernel is considered 'cheating' according to the following rules: \\

  1. **All Core Computation Logic Must Be Written in Triton** \\
    - All core computation logic must be implemented in Triton and must not delegate any part of the computation to high-level PyTorch APIs, such as torch.nn or torch.nn.functional. \\
    - Wrapping high-level torch APIs in a Triton kernel wrapper without expressing the computation logic in Triton is not acceptable. \\
    - Examples: \\
      (1). When implementing matrix multiplication in Triton, the use of \texttt{torch.matmul}, \texttt{@}, \texttt{torch.nn.Linear}, or similar high-order APIs is forbidden. \\
      (2). When implementing layer normalization in Triton, the use of \texttt{torch.nn.functional.layer\_norm} is forbidden. \\
      (3). When implementing softmax in Triton, the use of \texttt{torch.softmax} is forbidden. \\
      (4). When implementing convolution or convolution transpose, the use of \texttt{torch.nn.functional.conv2d} or \texttt{torch.nn.ConvTranspose3d} is forbidden. \\
      (5). When implementing attentions in Triton, the use of transformer library or other high-level APIs is forbidden. \\

  2. **No Trivial Identity or No-Op Computation** \\
    - Any part of the Triton kernel should always apply a mathematically-nontrivial operation to the input tensors like computation, reduction, or data layout change. \\
    - Examples: \\
      (1). Kernels that merely copy input to output (i.e., 'output\_val = input\_val') without performing meaningful computation are not allowed. \\
      (2). Kernels that does 'output\_val = input\_val + 0' or '* 1.0' are not allowed. \\

  3. **No Omission of Necessary Computations** \\
    - When implementing an operator that involves multiple components, the kernel must implement **all** required parts. Skipping any required mathematical component of the algorithm is considered cheating. \\
    - Examples: \\
      1. When implementing LayerNorm, the kernel must include computations of \\
      - Mean and variance computation \\
      - Normalization \\
      - Scaling and bias if applicable \\
      2. When implementing matmul followed by a transpose, the kernel must not omit the transpose of the matmul results. \\
      
  4. **The Kernel Must Address a Known Performance Bottleneck** \\
    - The operation should be **non-trivial** enough that writing a custom kernel could yield meaningful performance benefits using PyTorch API \\
    - Examples: \\
      - Fused element-wise ops (e.g., GELU + dropout + bias) \\
      - Reductions (e.g., softmax, LayerNorm, RMSNorm) \\
      - Operations that can benefit from different memory layout of intermediate tensors \\

  5. **The kernel Should Make Efficient Use of Available Parallelism in GPU** \\
    - Kernels should make use of block/thread-level parallelism ('program\_id', 'arange', 'BLOCK\_SIZE', etc.) \\
    - No scalar-only computations or logic that doesn’t scale with input size \\
\end{tcolorbox}

\section{Evaluation and examples}
\label{sec:appendix:eval}
\subsection{Evaluation with KernelBench}
\label{sec:kernelbench_example}
To evaluate the trained models, we construct prompts for 250 tasks in KernelBench. Similar to KernelBook, KernelBench provides a reference PyTorch code for each task.
For each given reference PyTorch code, we construct a prompt with one simple example pair of (PyTorch code, Triton kernel code), similarly to the one-shot prompting format in KernelBench.
Here, we use the following PyTorch and Triton codes for a simple add operation as an example:
\begin{lstlisting}[
  language=Python,
  basicstyle=\ttfamily\footnotesize,
  frame=single,
  breaklines=true,
  columns=fullflexible
]
### PyTorch reference code ###
import torch
import torch.nn as nn
import torch.nn.functional as F

class Model(nn.Module):
    def __init__(self) -> None:
        super().__init__()

    def forward(self, a, b):
        return a + b

def get_inputs():
    # randomly generate input tensors based on the model architecture
    a = torch.randn(1, 128).cuda()
    b = torch.randn(1, 128).cuda()
    return [a, b]

def get_init_inputs():
    # randomly generate tensors required for initialization based on the model architecture
    return []

### Triton kernel code ###

import torch
import torch.nn as nn
import torch.nn.functional as F
import triton
import triton.language as tl

@triton.jit
def add_kernel(
    x_ptr,  # Pointer to first input
    y_ptr,  # Pointer to second input
    out_ptr,  # Pointer to output
    n_elements,  # Total number of elements in input/output
    BLOCK_SIZE: tl.constexpr,
):
    # Each program handles a contiguous block of data of size BLOCK_SIZE
    block_start = tl.program_id(0) * BLOCK_SIZE
    # Create a range of offsets [0..BLOCK_SIZE-1]
    offsets = block_start + tl.arange(0, BLOCK_SIZE)
    # Mask to ensure we don't go out of bounds
    mask = offsets < n_elements
    # Load input values
    x = tl.load(x_ptr + offsets, mask=mask, other=0.0)
    y = tl.load(y_ptr + offsets, mask=mask, other=0.0)
    # Perform the elementwise addition
    out = x + y
    # Store the result
    tl.store(out_ptr + offsets, out, mask=mask)

def triton_add(x: torch.Tensor, y: torch.Tensor):
    """
    This function wraps the Triton kernel call. It:
      1. Ensures the inputs are contiguous on GPU.
      2. Calculates the grid (blocks) needed.
      3. Launches the Triton kernel.
    """
    assert x.is_cuda and y.is_cuda, "Tensors must be on CUDA."
    x = x.contiguous()
    y = y.contiguous()

    # Prepare output tensor
    out = torch.empty_like(x)

    # Number of elements in the tensor
    n_elements = x.numel()
    BLOCK_SIZE = 128  # Tunable parameter for block size

    # Determine the number of blocks needed
    grid = lambda meta: ((n_elements + meta["BLOCK_SIZE"] - 1) // meta["BLOCK_SIZE"],)

    # Launch the Triton kernel
    add_kernel[grid](x, y, out, n_elements, BLOCK_SIZE=BLOCK_SIZE)
    return out

class ModelNew(nn.Module):
    def __init__(self) -> None:
        super().__init__()

    def forward(self, a, b):
        # Instead of "return a + b", call our Triton-based addition
        return triton_add(a, b)
    
\end{lstlisting}

\begin{tcolorbox}[colback=gray!10, colframe=gray!80, title=Prompt (input) example]
You write custom Triton kernels to replace the pytorch operators in the given architecture to get speedups.

You have complete freedom to choose the set of operators you want to replace. You may make the decision to replace some operators with custom Triton kernels and leave others unchanged. You may replace multiple operators with custom implementations, consider operator fusion opportunities (combining multiple operators into a single kernel, for example, combining matmul+relu), or algorithmic changes (such as online softmax). You are only limited by your imagination.

Here's an example to show you the syntax of inline embedding custom Triton kernels in torch. The example given architecture is:

\texttt{``` <Example PyTorch code> ```}

The example new architecture with custom Triton kernels looks like this:

\texttt{``` <Example Triton kernel code> ```}

You are given the following architecture:

\texttt{``` <PyTorch reference code> ```}

Optimize the architecture named Model with custom Triton kernels! Name your optimized output architecture ModelNew. Output the new code in codeblocks. Please generate real code, NOT pseudocode, make sure the code compiles and is fully functional. Just output the new model code, no other text, and NO testing code!
\end{tcolorbox}

\subsection{Example of Triton codes generated by \ours (D) with speedup $>1$}
\label{sec:ex_kernel}

\textbf{Problem 12 in KernelBench Level 1: diagonal matrix multiplication with x11 speedup.}
\begin{lstlisting}[
  language=Python,
  basicstyle=\ttfamily\footnotesize,
  frame=single,
  breaklines=true,
  columns=fullflexible
]
import torch
import triton
import triton.language as tl

@triton.jit
def fused_diag_matmul_kernel(
  vec_a_ptr,      # Pointer to the diagonal vector A (N,)
  mat_b_ptr,      # Pointer to the input dense matrix B (N, M)
  output_ptr,     # Pointer to the output matrix (N, M)
  # --- Matrix dimensions ---
  N,              # The size of the diagonal vector A
  M,              # The number of columns in matrix B
  TOTAL_ELEMENTS, # Total number of elements in the output matrix (N * M)
  # --- Tuning parameters ---
  BLOCK_SIZE: tl.constexpr, # Number of elements each Triton program will handle
):
  """
  Computes C = diag(A) * B, where A is a vector representing the diagonal.
  This kernel treats the operation as a simple element-wise multiplication
  where each row of B is scaled by the corresponding element of A.
  """
  # 1. CALCULATE OFFSETS: Determine which elements this program instance will process.
  pid = tl.program_id(axis=0)
  block_start_offset = pid * BLOCK_SIZE
  offsets = block_start_offset + tl.arange(0, BLOCK_SIZE)

  # 2. CREATE BOUNDARY-CHECK MASK: Prevent out-of-bounds memory access.
  mask = offsets < TOTAL_ELEMENTS

  # 3. COMPUTE 2D INDICES FROM 1D OFFSETS: Find the row for vector A.
  row_indices = offsets // M

  # 4. LOAD DATA EFFICIENTLY: Read from global memory into registers.
  vec_a_elements = tl.load(vec_a_ptr + row_indices, mask=mask, other=0.0)
  mat_b_elements = tl.load(mat_b_ptr + offsets, mask=mask, other=0.0)

  # 5. PERFORM THE CORE COMPUTATION: The actual fused operation.
  output = vec_a_elements * mat_b_elements

  # 6. STORE THE RESULT: Write the computed values back to global memory.
  tl.store(output_ptr + offsets, output, mask=mask)
\end{lstlisting}

\textbf{Problem 94 in KernelBench Level 1: MSE (Mean Squared Error) with x3 speedup.}
\begin{lstlisting}[
  language=Python,
  basicstyle=\ttfamily\footnotesize,
  frame=single,
  breaklines=true,
  columns=fullflexible
]
import torch
import torch.nn as nn
import triton
import triton.language as tl

@triton.jit
def mse_kernel(
    pred_ptr,
    target_ptr,
    output_ptr,
    n_elements,
    BLOCK_SIZE: tl.constexpr,
):
    # 1. Block Setup
    pid = tl.program_id(axis=0)
    block_start = pid * BLOCK_SIZE
    offsets = block_start + tl.arange(0, BLOCK_SIZE)
    mask = offsets < n_elements

    # 2. Masked Load with Zero-Padding
    # CRITICAL: other=0.0 ensures that elements outside the valid range 
    # contribute 0 to the sum, preventing calculation errors.
    pred_vals = tl.load(pred_ptr + offsets, mask=mask, other=0.0)
    target_vals = tl.load(target_ptr + offsets, mask=mask, other=0.0)

    # 3. Compute Squared Error
    # This happens in parallel across all threads in the block
    diff = pred_vals - target_vals
    sq_diff = diff * diff

    # 4. Intra-Block Reduction
    # tl.sum collapses the 1024 values in this block into a single scalar.
    # This is extremely fast as it uses shared memory/registers, avoiding VRAM.
    block_sum = tl.sum(sq_diff, axis=0)

    # 5. Global Accumulation
    # We use an atomic add to aggregate the partial sums from each block 
    # into the single global output location.
    tl.atomic_add(output_ptr, block_sum)

def triton_mse(pred: torch.Tensor, target: torch.Tensor) -> torch.Tensor:
    # Ensure data layout is contiguous for coalesced memory access
    if not pred.is_contiguous(): pred = pred.contiguous()
    if not target.is_contiguous(): target = target.contiguous()
        
    n_elements = pred.numel()
    if n_elements == 0:
        return torch.tensor(0.0, device=pred.device)

    # Output must be initialized to zero because atomic_add accumulates on top of it
    output = torch.zeros(1, device=pred.device, dtype=pred.dtype)
    
    BLOCK_SIZE = 1024
    # Standard grid calculation
    grid = lambda meta: (triton.cdiv(n_elements, meta['BLOCK_SIZE']),)
    
    mse_kernel[grid](pred, target, output, n_elements, BLOCK_SIZE=BLOCK_SIZE)
    
    # Final division by N happens on Python side (negligible cost)
    return output[0] / n_elements

class ModelNew(nn.Module):
    def forward(self, predictions, targets):
        return triton_mse(predictions, targets)
\end{lstlisting}

\textbf{Problem 99 in KernelBench Level 1: Triplet Margin Loss with x4 speedup.}
\begin{lstlisting}[
  language=Python,
  basicstyle=\ttfamily\footnotesize,
  frame=single,
  breaklines=true,
  columns=fullflexible
]
import torch
import torch.nn as nn
import triton
import triton.language as tl

@triton.jit
def triplet_margin_loss_kernel(
    anchor_ptr,
    positive_ptr,
    negative_ptr,
    output_ptr,
    margin,
    # FEATURE_SIZE as constexpr allows the compiler to optimize the loop unrolling
    FEATURE_SIZE: tl.constexpr,
    BLOCK_SIZE: tl.constexpr,
):
    # 1. Parallelize by Batch (Row)
    # Each program instance (block) handles exactly one triplet sample from the batch.
    pid = tl.program_id(axis=0)
    
    # Calculate starting memory addresses for this specific row
    offset_start = pid * FEATURE_SIZE
    anchor_row = anchor_ptr + offset_start
    positive_row = positive_ptr + offset_start
    negative_row = negative_ptr + offset_start
    
    # Registers to accumulate squared distances across the loop
    ap_sum = 0.0
    an_sum = 0.0
    
    # 2. Tiled Processing of Features
    # We loop over the feature dimension in chunks (tiles) of BLOCK_SIZE.
    # This keeps register pressure low and allows handling features > 1024 dims.
    for j in range(0, FEATURE_SIZE, BLOCK_SIZE):
        # Generate offsets for the current tile
        offsets = j + tl.arange(0, BLOCK_SIZE)
        mask = offsets < FEATURE_SIZE
        
        # Load data chunks (coalesced reads)
        a = tl.load(anchor_row + offsets, mask=mask, other=0.0)
        p = tl.load(positive_row + offsets, mask=mask, other=0.0)
        n = tl.load(negative_row + offsets, mask=mask, other=0.0)
        
        # Vectorized math on the chunk
        diff_ap = a - p
        sq_ap = diff_ap * diff_ap
        
        diff_an = a - n
        sq_an = diff_an * diff_an
        
        # 3. Partial Reduction
        # Reduce the chunk to a scalar and add to the running total.
        # tl.sum is highly optimized for intra-block communication.
        ap_sum += tl.sum(sq_ap, axis=0)
        an_sum += tl.sum(sq_an, axis=0)
        
    # 4. Final Loss Calculation (per row)
    d_ap = tl.sqrt(ap_sum)
    d_an = tl.sqrt(an_sum)
    
    loss_i = d_ap - d_an + margin
    loss_i = tl.where(loss_i > 0, loss_i, 0.0) # ReLU
    
    # 5. Global Accumulation
    # Aggregate this row's loss into the total batch loss.
    # Note: For massive batch sizes, this atomic_add can become a contention hotspot.
    tl.atomic_add(output_ptr, loss_i)

class ModelNew(nn.Module):
    def __init__(self, margin=1.0):
        super().__init__()
        self.margin = margin
        
    def forward(self, anchor, positive, negative):
        # Ensure contiguous memory for safe pointer arithmetic
        anchor = anchor.contiguous()
        positive = positive.contiguous()
        negative = negative.contiguous()
        
        batch_size, n_features = anchor.shape
        
        # Allocate output tensor (scalar)
        total_loss = torch.zeros(1, device=anchor.device, dtype=anchor.dtype)
        
        # Grid Size = Batch Size (one block per sample)
        grid = (batch_size,)
        
        triplet_margin_loss_kernel[grid](
            anchor,
            positive,
            negative,
            total_loss,
            self.margin,
            FEATURE_SIZE=n_features, # Pass actual dimension dynamically
            BLOCK_SIZE=1024,         # 1024 is standard for max occupancy
        )
        
        return total_loss[0] / batch_size
\end{lstlisting}

\end{document}